\documentclass[iop]{emulateapj}
\usepackage{amssymb}
\usepackage{natbib}
\usepackage{amsmath}
\usepackage{subfigure}
\usepackage{footnote}
\usepackage{color}
\usepackage{mdwlist}
\usepackage{enumerate}

\newcommand{\xiv}{{\ion{Fe}{14}}}
\newcommand{\xvi}{{\ion{Fe}{16}}}
\newcommand{\xxi}{{\ion{Fe}{21}}}
\newcommand{\xxiii}{{\ion{Fe}{23}}}
\newcommand{\xxiv}{{\ion{Fe}{24}}}
\newcommand{\xix}{{\ion{Fe}{19}}}
\newcommand{\cii}{{\ion{C}{2}}}

\newcommand{\ci}{{\ion{C}{1}}}
\newcommand{\oi}{{\ion{O}{1}}}
\newcommand{\mgii}{{\ion{Mg}{2}}}
\newcommand{\siiv}{{\ion{Si}{4}}}

\newcommand{\sii}{{\ion{S}{2}}}
\newcommand{\feii}{{\ion{Fe}{2}}}
\newcommand{\oiv}{{\ion{O}{4}}}
\newcommand{\sivii}{{\ion{Si}{7}}}
\newcommand{\xviii}{{\ion{Fe}{18}}}
\newcommand{\si}{\ion{S}{1}}
\newcommand{\siv}{{\ion{S}{4}}}

\begin{document}

\title{Simultaneous IRIS and Hinode/EIS observations and modelling of the 27 October 2014 X~2.0~class flare}
\author{V.Polito}
\affil{DAMTP, CMS, University of Cambridge, Wilberforce Road, Cambridge CB3 0WA, United Kingdom}

\author{J. W. Reep}
\affil{National Research Council Post-Doc Program, Naval Research Laboratory, Washington, DC 20375, USA}
\affil{formerly DAMTP, CMS, University of Cambridge, Wilberforce Road, Cambridge CB3 0WA, United Kingdom}
\author{K.K. Reeves}
\affil{Harvard-Smithsonian Center for Astrophysics, 60 Garden Street, Cambridge MA 01238, USA}
\author{P. J. A. Sim\~oes}
\affil{SUPA School of Physics and Astronomy, University of Glasgow, Glasgow G12
8QQ, UK}
\author{J.Dud\'{i}k}
\affil{Astronomical Institute, Academy of Sciences of the Czech Republic, 25165 Ond\v{r}ejov, Czech Republic}
\author{G. Del Zanna}
\author{H.E. Mason}
\affil{DAMTP, CMS, University of Cambridge, Wilberforce Road, Cambridge CB3 0WA, United Kingdom}
\author{L. Golub}
\affil{Harvard-Smithsonian Center for Astrophysics, 60 Garden Street, Cambridge MA 01238, USA}

\begin{abstract}
We present the study of the X2-class flare which occurred on the 27 October 2014 and was observed with the Interface Region Imaging Spectrograph (IRIS) and the EUV Imaging Spectrometer (EIS) on board the Hinode satellite. Thanks to the high cadence and spatial resolution of the IRIS and EIS instruments, we are able to compare simultaneous observations of the \xxi~1354.08~\AA~and \xxiii~263.77~\AA~high temperature emission ($\gtrsim$ 10~MK) in the flare ribbon during the chromospheric evaporation phase. We find that IRIS observes completely blue-shifted \xxi~line profiles, up to 200 km s$^{-1}$ during the rise phase of the flare, indicating that the site of the plasma upflows is resolved by IRIS. In contrast, the \xxiii~line is often asymmetric, which we interpret as being due to the lower spatial resolution of EIS. Temperature estimates from SDO/AIA and Hinode/XRT show that hot emission (log($T$)[K] $>$ 7.2) is first concentrated at the footpoints before filling the loops. Density sensitive lines from IRIS and EIS give electron number density estimates of $\gtrsim$~10$^{12}$~cm$^{-3}$ in the transition region lines and 10$^{10}$~cm$^{-3}$ in the coronal lines during the impulsive phase. In order to compare the observational results against theoretical predictions, we have run a simulation of a flare loop undergoing heating using the HYDRAD 1D hydro code. We find that the simulated plasma parameters are close to the observed values which are obtained with IRIS, Hinode and AIA. These results support an electron beam heating model rather than a purely thermal conduction model as the driving mechanism for this flare.
\end{abstract}

\section{Introduction }
Finding a definitive model to explain the physical processes involved in flares still represents one of the most important challenges in solar physics. For four decades, there has been considerable interest and effort placed in numerical modelling of solar flares. 

The thermal conduction model, where flare energy is deposited at the apex of a loop to drive a thermal conduction front, has been investigated numerically by several authors, e.g.~\citet{nagai1980}, ~\citet{nagai1983},~\citet{cheng1983},~\citet{pallavicini1983},~\citet{macneice1986},~\citet{tsiklauri2004}, and~\citet{bradshaw2004}. There has been
similar interest in the thick-target model, wherein a beam of energetic electrons stream from the acceleration region above the loop-top towards the chromosphere where they deposit their energy through Coulomb collisions with the ambient plasma.  A number of authors have investigated this model extensively:~\citet{nagai1984},\citet{macneice1984},~\citet{fisher1985a},~\citet{fisher1985b},~\citet{mariska1989},~\citet{abbett1999}, ~\citet{allred2005},~\citet{cheng2010},~\citet{Winter2011},~\citet{reep2013},~\citet{kowalski2015} and~\citet{reep2015}.  There has also been interest in the possibility of Alfvenic wave heating as a driving mechanism of solar flares \citep{emslie1982, Fletcher08,russell2013,melrose2014}. 

Regardless of the energy deposition mechanism at play, the result is an intense heating and overpressure of the chromospheric plasma at the flare \emph{kernels}, resulting in a consequent evaporation (\emph{chromospheric evaporation}) and filling of the flare loops, which become visible in the Extreme Ultra Violet (EUV) and soft X-ray (SXR) images.  Even though different theoretical models have had varying success in reproducing and explaining some observational features, we still cannot fully explain the complex nature of a large number of flares studied over the past decades \citep[see, e.g., an observational review by][]{Fletcher11}. 

One the major problems is related to the fact that we are not able to directly observe the site of the heating release and the details of the energy conversion. EUV and X-ray spectroscopy provide a powerful tool to study the response of the chromospheric plasma to the heating and to test the flare loop models. 
In particular, 1D hydrodynamics models \citep[e.g.,][]{Emslie87} predict completely blueshifted EUV and SXR profiles as a result of chromospheric evaporation during the impulsive phase of the flare. In addition, emission lines formed at higher temperatures are predicted to show higher evaporation velocities.

Large blue shifts in the high temperature lines were first observed in the soft X-ray wavelength range (8-25 MK) with SOLFLEX \citep{Doschek79} and the Solar Maximum Mission (SMM) \citep{Antonucci82}. The observed line profiles were dominated by a strong rest component superimposed with a blue wing of up to hundreds of km\,s$^{-1}$, contrary to the theoretical prediction of completely blue-shifted profiles. One possible interpretation was that the instruments observed a superposition of a stationary component from the flare loops and a blue-shifted component from the kernels. Asymmetric profiles with blue wing enhancements at around $\sim$ 200~km\,s$^{-1}$ were also measured with the SMM Ultraviolet Spectrometer and Polarimeter in the lower temperature \xxi~1354.08~\AA~line ($T$ $\gtrsim$~10~MK) \citep{Mason86}. However, the X-ray studies lacked good spatial information and it was therefore difficult to give a straightforward interpretation of the observed spectra in terms of theoretical models. 

The launch of the Coronal Diagnostic Spectrometer \citep[CDS;][]{Harrison95} allowed spatially resolved spectroscopy (4--5$\arcsec$) over a range of temperatures, including the high temperature \xix~(9 MK) flare line. 
 For instance,~\citet{Teriaca03},~\citet{Brosius03},~\citet{Milligan06} observed \xix~asymmetric profiles with a dominant high velocity component at $\sim$~200~km\,s$^{-1}$. In addition, \citet{DelZanna06} reported the observation of a completely blue-shifted ($\sim$~140~km\,s$^{-1}$) \xix~profile during the impulsive phase of an M1 class flare and small redshifts in the cooler lines, in agreement with the theoretical predictions that emission lines formed at higher temperature have larger velocities. The CDS results suggested that the evaporation sites were close to being spatially resolved. It was still unclear, however, why in some instances CDS observed asymmetric and not entirely blue-shifted line profiles.
 
The EUV Imaging Spectrometer \citep[EIS;][]{Culhane97} on board the Hinode satellite launched in 2006, extended the CDS capabilities allowing observations of the evaporation site over a broader range of temperatures with higher spatial ($\sim$ 3$\arcsec$) resolution \citep[e.g.,][]{Milligan09,Watanabe10,DelZanna11,Graham11, Young13, Doschek13}. In particular, \citet{Watanabe10} and \citet{Young13} observed dominant blue-shifted components of the order of 400~km\,s$^{-1}$ in the EUV \xxiii~and \xxiv~lines (formed at $\gtrsim$~10~MK). Moreover, completely blue-shifted \xxiii~emission ($\sim$~200~km\,s$^{-1}$) was observed by \citet{Brosius13} for a C1 class flare.

Despite the high spatial resolution of the CDS and EIS spectrometers, some of the observations showed high temperature line profiles which could be fitted with multi Gaussian components. A possible interpretation is that the observed emission was the result of a superposition of different plasma upflows along the line-of-sight \citep{Doschek05, Reeves2007}. However, the exact origin of the stationary component remained uncertain. 

With the advent of the Interface Region Imaging Spectrograph \citep[IRIS;][]{DePontieu14}, the dynamics of the flaring plasma can now be investigated with unprecedented spatial resolution, (0.33--0.4$\arcsec$), a cadence (up to 2s) and a high spectral resolution allowing the determination of accurate velocities($\sim$~3~km\,s$^{-1}$). The IRIS spectrograph (SP) observes continua and emission lines over a very broad range of temperatures (log($T$[K])~=~3.7--7), including the flare line \xxi~1354.08~\AA~formed at $\sim$~11~MK. 
Simultaneously, the IRIS Slit Jaw Imager (SJI) provides high-resolution context images in four different passbands (\cii~1330~\AA~, \siiv~1400~\AA, \mgii~k 2796~\AA~ and \mgii~wing 2830~\AA~). Since its launch in 2013, IRIS has observed several flare events and opened up new prospects for the investigation of solar flare dynamics. One interesting feature is the absence of \xxi~multi-components profiles including a rest wavelength component for some large flares observed by IRIS. In particular, blue-shifted \xxi~profiles of up to 200-300 km\,s$^{-1}$ have been reported by~\citet{Young15},~\citet{Tian15}, and \citet{Graham15} during two X-class flares observed in 2014. In these studies, the \xxi~line was always entirely blue-shifted during the impulsive phase, confirming the CDS and EIS results of e.g.,~\citet{DelZanna06} and \citet{DelZanna11}.~In addition, in the study of a small C-class flare,~\citet{Polito15} observed completely blue-shifted ($\sim$ 80 km\,s$^{-1}$) \xxi~symmetric profiles which became asymmetric and dominated by a rest emission $\sim$ 50 s after the evaporation first occurred. This rest emission was interpreted as originating from the overlying flare loops along the line of sight. Interestingly, the IRIS \xxi~blueshifts seem to be significantly lower than the \xxiii~and \xxiv~Doppler shifts for some of the flare observations with EIS. 

Simultaneous flare studies with EIS and IRIS hence offer the unique opportunity to compare simultaneous observations of \xxi,~\xxiii~and \xxiv~blueshifts in order to clarify if there are indeed differences in emission lines formed over similar temperatures during the chromospheric evaporation phase and if the instrumental resolution plays a role. 

Joint spectroscopic observations present complications associated with the co-alignment of the EIS and IRIS slits rastering in different directions over a small field of view. Using multi wavelength imaging is therefore crucial in order to understand the context of the observed events. Since 2010, the Atmospheric Imaging Assembly \citep[AIA;][]{Lemen12} on board the Solar Dynamics Observatory (SDO) has provided continuous (12 s cadence) multi-wavelength high resolution (1.2$\arcsec$ $\sim$ 800 km) images of the full Sun in 8 EUV and 2 UV filters. In particular, the two 131~\AA~and 94~\AA~passbands represent a powerful tool to diagnose the high temperature plasma during flares \citep[see for example,][]{Petkaki12}.

 We performed a detailed search for flare events where both IRIS and EIS were observing the flare footpoints during the impulsive phase. The event we selected for further investigation was an X2.0 class flare which occurred in October 2014. Our aim is to (1) compare simultaneous observations with EIS and IRIS of the \xxiii~and \xxi~emission during the chromospheric evaporation phase, (2) compare plasma parameters derived from diagnostics with IRIS, Hinode and AIA  observations to a detailed theoretical model using the HYDRAD code \citep{bradshaw2003,bradshaw2013,reep2013}. Data from the Ramaty High Energy Solar Spectroscopic Imager \citep[RHESSI;][]{Lin:2002} and the Geostationary Operational Environment Satellite (GOES) were also used to provide information about the electron beam source and energy deposition rate, in terms of a collisional thick-target flare model.
 
The paper is structured as follows: Sect. \ref{Sect.:2} presents the context of the AIA, IRIS, EIS and RHESSI observations. The general evolution of the flare is then discussed in Sect. \ref{Sect.:3}, while Sect. \ref{Sect.:4} presents the details of the chromospheric evaporation observed by IRIS and EIS. The main results of the plasma diagnostics performed with IRIS, EIS and AIA observations is then reported in Sect. \ref{Sect.:5}. The hydrodynamic simulations with the HYDRAD code and the comparison with the observational results are discussed in Sect. \ref{Sect.:6}. Finally, in Sect. \ref{Sect.:7} we discuss and summarize our work.
\section{Observation of the X2 class flare on the 27~October~2014}
\label{Sect.:2}
\begin{figure}[!ht]
	\centering
	\includegraphics[width=0.5\textwidth]{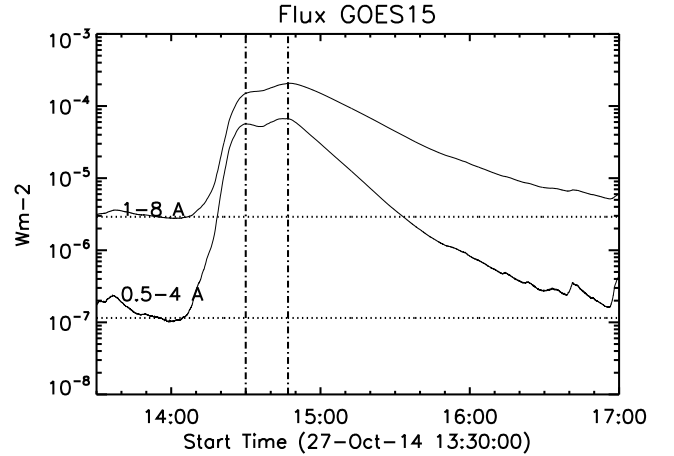} 	
      \caption{Light curves of the 27 October 2014 flare in the two GOES channels: 0.5--4 \AA~ and 1--8 \AA. We observe two peaks in the soft X-ray signal (indicated by dash-dot lines) suggesting the occurrence of two distinct heating events for the flare under study.}
      \label{Fig:goes}
  \end{figure}
\begin{figure*}[!htl]
	\centering
	\includegraphics[width=\textwidth]{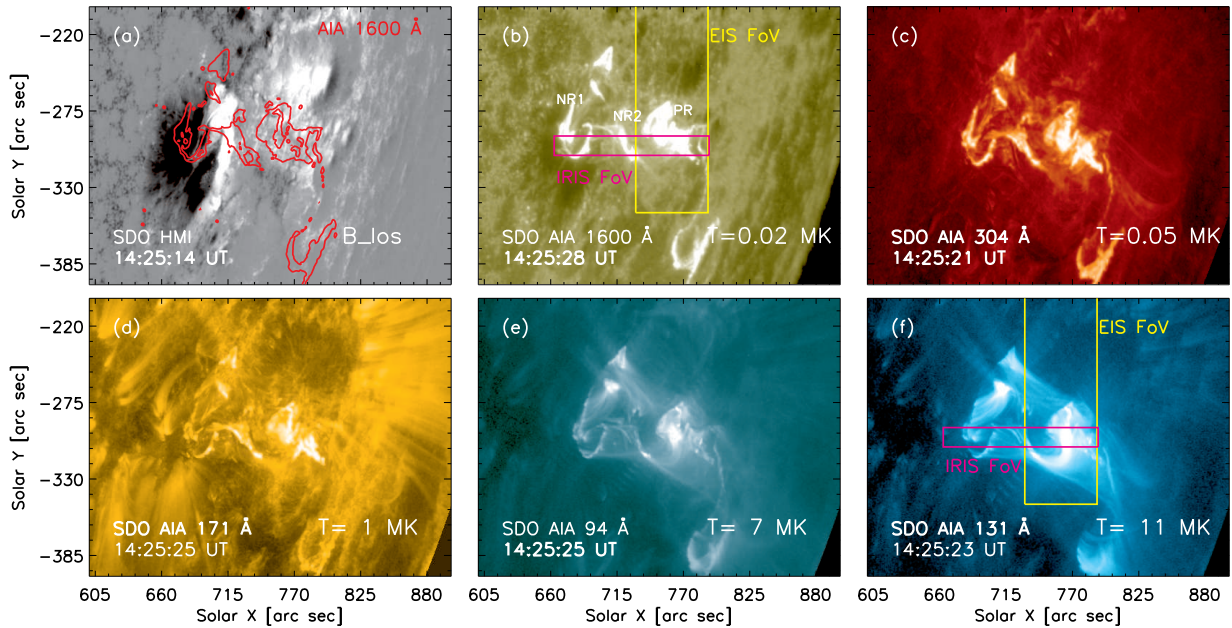} 	
      \caption{HMI los B (a), AIA 1600~\AA~ (b), 304~\AA~(c), 171~\AA~(d), 94~\AA~(e) and 131~\AA~(f) frames showing the 27 October 2014 X class flare at around 14:25~UT, close to the first flare peak observed in the soft X-ray with GOES. The pink and yellow boxes overlaid on the panels (b) and (f) represent the IRIS and EIS fields-of-view, respectively. See also the online Movie 1 showing the evolution of the HMI los B and AIA images over time.}
      \label{Fig:context}
  \end{figure*}  
The Active Region (hereafter, AR) NOAA 12192 was a large and strongly flaring $\beta\gamma\delta-\beta\gamma\delta$ active region during its entire disk passage on 2014 October 18--30. It consisted of a leading positive-polarity sunspot and two large negative-polarity following sunspots, and multiple plage polarities. The sunspots contained multiple opposite-polarity intrusions. It produced 55 C-class, 23 M-class, and 4 X-class flares. The M- and X-class flares occurred in the AR 12192 were investigated by \citet{Chen2015} using SDO/AIA and the Helioseismic and Magnetic Imager \citep[HMI;][]{Scherrer2012} data. In this study, we focus on the X2.0 class flare that occurred on the 27 October 2014 from around 14:00 UT to 16:31 UT (see Fig. \ref{Fig:context}), as measured by the GOES-15 satellite in the 1--8 \AA~and 0.5--4 \AA~channels (Fig. \ref{Fig:goes}). The GOES soft X-ray light curves exhibit two close peaks at around 14:30 and 14:47~UT, which suggest two heating events in the AR 12192 for the flare under study. 

\subsection{Set of observations and data reduction}

\subsubsection{SDO/AIA and HMI data}
The SDO/AIA and HMI level 1 data were downloaded  using the IDL routines included in the solarsoft \emph{VSO} package and converted to level 1.5 images with the \emph{aia\_prep.pro} and \emph{hmi\_prep.pro} routines. The images were also corrected for solar rotation. 

Each of the 10 EUV/UV AIA filters include several strong emission lines which are formed at different plasma temperatures, resulting in a multi-thermal response of the instrument \citep[e.g.,][]{ODwyer10,DelZanna11b}. It is therefore essential to compare AIA images in different filters to fully understand the context of the observation. 
Fig. \ref{Fig:context} shows SDO/AIA images of the flare in the 1600~\AA~(b), 304~\AA~(c), 171~\AA~(d), 94~\AA~(e) and 131~\AA~(f) channels. In addition, a line of sight magnetic field ($B_\mathrm{LOS}$) map taken by HMI is also shown in the panel (a). These images were taken close to the first peak of the GOES X-ray curve (Fig. \ref{Fig:goes}), at around 14:27~UT. The flare negative and positive polarity ribbons are best seen in the 1600~\AA~and 304~\AA~images, and the 1600~\AA~contours have been overlaid on the magnetic field image. The positive and main negative polarity ribbons are indicated by PR and NR1 respectively in the panel (b) of Fig. \ref{Fig:context}. A secondary negative ribbon NR2 develops during the evolution of the flare as explained in Sect. \ref{Sect.:3}. We can observe from the panel (b) in Fig. \ref{Fig:context} that NR2 seems to span both negative and positive polarities. Close to the limb, strong magnetic polarities can in fact exhibit a shadow of opposite polarity, which is due to the projection effects \citep{Venkatakrishnan88,Gary90} caused by the mis-match between the $B_\mathrm{LOS}$ and the locally vertical component of the magnetic field. The 131~\AA~images are dominated by plasma at T $\sim$~10~MK \citep{Petkaki12} and show the flare loops formed between the PR and NR2 ribbons. Finally, the fields of view of EIS and IRIS are indicated by the yellow and pink boxes respectively. 

The online Movie 1 shows the overview of the flare evolution as seen in the HMI ($B_\mathrm{LOS}$) maps and in the 5 AIA passbands of Fig.\ref{Fig:context}. 
 The AIA data were also used to co-align the spectroscopic images and provide powerful plasma diagnostics, as explained in sections \ref{Sect.:2.3}, \ref{Sect.:5.3} and \ref{Sect.:5.4}. 
 
 \begin{figure*}[!htl]
	\centering
	\includegraphics[scale=0.7]{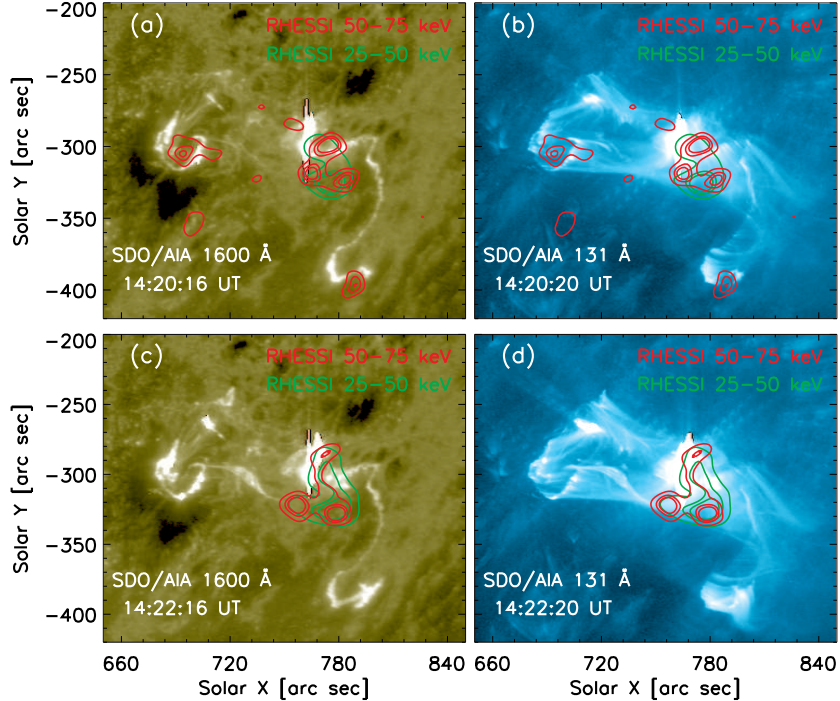} 	
      \caption{RHESSI HXR 25--50 keV and 50--75 keV intensity contours overlaid on AIA 1600~\AA~and 131~\AA~images during the impulsive phase of the flare. The contours show the 50, 70, 80 $\%$ of the maximum RHESSI HXR signal.}
      \label{Fig:rhessi}
  \end{figure*}
 
\subsubsection{IRIS}
On the 27 October 2014 IRIS was running a large coarse 8-steps raster on the AR 12192 over the period 14:04~UT to 17:44~UT. During each raster, the 0.33$\arcsec$ slit was scanning from east to west over a field of view of 14$\arcsec$ $\times$ 119$\arcsec$~taking 26 seconds. The single slit positions were separated by 2$\arcsec$~and had an exposure time of 3 s, resulting in an actual step cadence of about 3.2 s. The IRIS Slit Jaw Imager (SJI) obtained images every 12 s alternatively in the \cii~1330~\AA,~\mgii~2796~\AA~and 2832~\AA~filters over an area of 120$\arcsec$ $\times$ 119$\arcsec$ on the AR under study. We used IRIS level 2 data, obtained from level 0 after flat-field, geometry calibration and dark current subtraction. The removal of cosmic rays was performed on the spectrometer data by using the solarsoft routine $\emph{despik.pro}$.  
Nine spectral windows were present in this study, but we only focused on the FUVS \oi~1355.60~\AA~and FUVL \siiv~1402.77~\AA~windows. 
The wavelength scale of the IRIS spectrometer is subject to a drift of around 8 km\,s$^{-1}$ (for the FUV channel, IRIS TN20\footnotemark[1]) as a result of the temperature variation of the detectors during one satellite orbit. To correct for this effect, we measured the orbital variation of the \oi~centroid position over time and subtracted it from both the FUVS and FUVL wavelength arrays. The Doppler shift of this photospheric line  is usually less than 1 km\,s$^{-1}$ and therefore represents a suitable reference for wavelength calibration purposes. 
Moreover, an absolute calibration can be obtained by using the strongest photospheric lines present in each spectral window. For the FUVS channel, we estimated the difference between the \oi~line position after the orbital correction and the expected rest wavelength of 1355.598~\AA~\citep{Sandlin86}. For the FUVL CCD, the \si~1401.515~\AA~ line was used. 

During flares, the intensities of chromospheric lines are highly enhanced at the ribbons and can cause blends with the higher temperature lines. The most important blending on the red side of the \xxi~1354.08~\AA~is the chromospheric \ci~1354.29~\AA~line. In addition, usually weak spectral lines such as \feii~and \sii~are observed in the IRIS \oi~spectral window (1352.6~\AA-1354.5~\AA), as shown by \citet{Young15}, \citet{Polito15}, \citet{Tian15}, and \citet{Graham15}. The width of these lines is usually narrow (typically around 0.06~\AA, but can be broader during the flare), and their profiles can be easily de-blended from the very broad \xxi~emission in most of the spectra. The expected width of the \xxi~line is estimated around 0.43~\AA, given by the quadratic sum of the instrumental width \citep[0.026~\AA,][]{DePontieu14} and of the line thermal width assuming a peak temperature of ion abundance of $1.1 \cdot 10^{7}$~K from CHIANTI \citep{Dere:1997, DelZanna2015}. The \xxi~is however observed to be significantly larger during the impulsive phase of flares \citep{Mason86,Polito15}. The FUV continuum is also strongly enhanced at the flare ribbon, adding a significant uncertainty when measuring weak \xxi~line profiles. To reduce this effect, we averaged the \xxi~spectra over a few pixels along the solar-Y direction above the ribbon, where the FUV continuum and low temperature emission lines are less enhanced, as explained in Sect. \ref{Sect.:3}. The \oiv~line profiles in the FUVL \siiv~spectral window are also affected by line blends, as discussed in Sect. \ref{Sect.:5.1}. 

The online Movie 2 shows the IRIS SJI 1330~\AA~and AIA 131 \AA~images over time. A movie frame is shown in Fig. \ref{Fig:loop_14_19} of Sect. \ref{Sect.:4}.

\footnotetext[1]{http://iris.lmsal.com/documents.html}
\subsubsection{Hinode/EIS and XRT}
\label{Sect.:2.3}

The EIS spectrometer was running a \textit{HH\_Flare\_raster\_v6} sparse raster study from 11:02:37~UT to 17:27:48~UT on the 27 October 2014, scanning the AR 12192 from west-to-east over an area of 162$\arcsec$ $\times$ 152$\arcsec$ taking around 212 s. Each scan contained 20 of the  2'' slit steps with a 1'' jump between each position. The exposure time was 9 s. 

The EIS data have been reduced to level 1 with the Solarsoft IDL routine $eis\_prep.pro$. We used standard options in order to remove hot and dusty pixels and interpolate missing pixels, as described in the EIS notes~ \footnotemark[2]. We have then applied the radiometric calibration by using \citet{DelZanna13} method, which corrects for the degradation of the instrument efficiency over time. The data have also been corrected to account for the offset (about 18 pixels) in the solar-$Y$ direction between the LW and SW CCD channels. When performing velocity measurements with EIS, the main issue is represented by the lack of low temperature reference lines to perform an absolute calibration of the spectra. In this study, we mainly focus on the \xxiii~263.765~\AA~high temperature line, which is only visible at the flare location. In order to estimate the expected \xxiii~rest wavelength, we measured the centroid position of the neighbouring \xvi~263.984~\AA~line in a background region outside the flare. The difference between the observed and the expected \xvi~wavelength thus provides a good estimate of the spectral drift of the \xxiii~spectrum. We acquired a reference \xvi~line for each EIS raster analyzed in this study, in order to remove the effect of the periodic shift of the wavelength scale during the orbital motion of the satellite. 

\footnotetext[2]{http://solarb.mssl.ucl.ac.uk:8080/eiswiki/}
The expected \xxiii~line width is $\sim$~0.12~\AA, given by the quadratic sum of the line thermal width and the EIS instrumental width. The thermal width is determined by assuming an isothermal temperature of 1.4~$\cdot$~10$^{7}$~K for the emitting plasma, while the instrumental width is given by the \textit{eis\_slit\_width.pro} solarsoft routine.

SDO/AIA and SJI images were used to precisely co-align the monochromatic images from EIS and IRIS. In particular, EIS \xvi~262.98~\AA~images can be aligned to AIA 335~\AA~maps, which are mainly dominated by \xvi~emission, formed at about $T \sim$ 3~MK. The IRIS SJI 1330~\AA~images mainly includes emission from \cii~1334/1335\AA~formed at $T \sim$ 0.02~MK and can thus be directly compared and aligned with AIA 1600~\AA~chromospheric images. Once the EIS and IRIS images are both aligned to co-temporal AIA images, we can then derive the relative co-alignement between the two spectrometers. All the images were cross-aligned manually, giving an uncertainty of about 2 AIA pixels ($\sim$ 1.2 arcseconds). Taking into account the EIS point-spread-function of 3$\arcsec$, the total alignment uncertainty between IRIS and EIS can be estimated to be around 3--5$\arcsec$.

The Hinode/X-Ray Telescope \citep[XRT;][]{Golub2007} performed measurements with few selected filters and at various times during the 27 October 2014 flare. In this study, we only use images in the Be\_thick and Al\_thick filters, which have a similar cadence during the flare. XRT level 0 data were converted to level 1 by using the solarsoft \textit{xrt\_prep.pro} routine, which subtracts the dark current and removes the CCD bias and telescope vignetting \citep{Kobelski2014}.
\subsubsection{RHESSI}
\label{Sect.:2.4}
We used the HXR data from RHESSI to obtain the location and spectral characteristics of the HXR emission. We constructed RHESSI HXR images using CLEAN \citep{Hurford:2002}, using front detectors 3 to 9, at 12--25, 25--50 and 50--75 keV throughout the impulsive phase, integrating in time bins of 20 seconds. In Fig. \ref{Fig:rhessi} we show the HXR sources at 25--50 and 50--75 keV, at the intervals 14:20:18--14:20:38~UT and 14:22:18--14:22:38~UT. The 12--25 keV source is co-spatial with the 25--50 keV, and thus it is not shown. The 25--50 keV source is associated well with the loops in the AIA 131~\AA~image, which connect the ribbons seen in the 1600~\AA~image. Fig. \ref{Fig:rhessi}a and \ref{Fig:rhessi}b show HXR sources associated with the positive and negative polarity ribbons (PR and NR1 respectively), as well as with the main flaring region. Near the maximum of the HXR emission (Fig. \ref{Fig:rhessi}c and \ref{Fig:rhessi}d) the emission is dominated by the main flaring region, with the northern and eastern sources probably associated with the ribbons, while the southern source is possibly a looptop source \citep[e.g.][]{Simoes:2013}, given the lack of chromospheric features in the IRIS SJI 1330~\AA~(see Fig. \ref{Fig:overview_131_SJI}) and AIA 1600~\AA. We believe that the co-alignment of AIA and RHESSI images is very good - see for instance the good agreement of the outermost HXR sources and the 1600~\AA~ ribbons in Fig. \ref{Fig:rhessi}. 
We analysed the HXR spectrum during the impulsive phase in order to estimate the properties and energy content of the accelerated electrons. This was performed using the standard software and procedures from OSPEX \citep{Schwartz:2002}. We fitted the HXR count spectrum assuming an isothermal plus thick-target model (\verb$thick2_vnorm$) to account for the thermal and non-thermal emission. The albedo effect was also considered in the analysis \citep{Kontar:2006}. Count spectra from RHESSI's front detectors 1, 3, 6 and 9 were fitted individually, then the results were averaged \citep[cf.][]{Milligan:2014}.
We estimated the power $P_\mathrm{nth}$ contained by the non-thermal electrons under the assumption of a thick-target model. Assuming that the electron distribution $F(E)$ has a power-law form $AE^{-\delta}$ (electrons s$^{-1}$ keV$^{-1}$), where $A$ is the normalisation factor (proportional to the total electron rate), $E_C$ is the low energy cutoff, and $\delta$ is the spectral index, a lower limit for the total power contained in the distribution can be estimated by \citep[see e.g.][]{Milligan:2014}:

\begin{equation}
P_\mathrm{nth}(E \ge E_C) = \int_{E_C}^\infty E F(E) dE \ \mathrm{erg s^{-1}}
\label{eq:power}
\end{equation}

\begin{figure}
  \centering
  \includegraphics[width=0.5\textwidth]{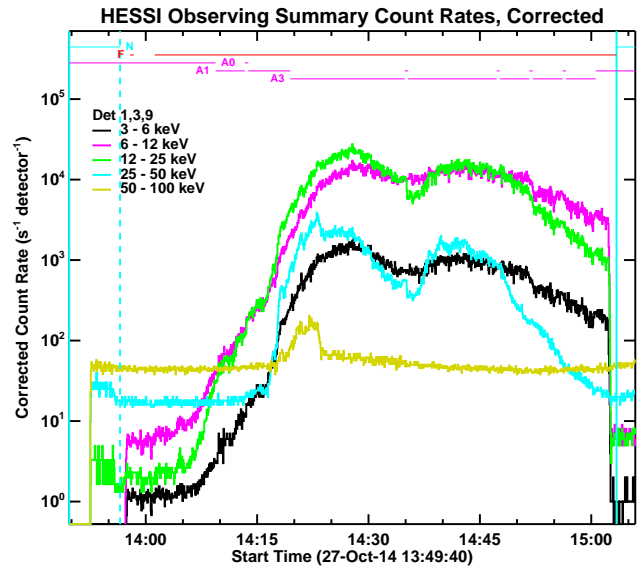}  
      \caption{RHESSI HXR corrected counts curves in the energy bands 3--6, 6--12, 12--25, 25--50 and 50--100 keV. RHESSI attenuator states (A0, A1 and A3) are show at the top of the figure.}
      \label{Fig:rhessi_obs}
  \end{figure}

The maximum power $P_\mathrm{nth}$ is $3.8 \pm 1.7 \times 10^{29}$ erg s$^{-1}$ around 14:27:06~UT, near the peak of the 12--25 keV curve (see Fig. \ref{Fig:rhessi_obs}).

\section{Evolution of the flare}
\label{Sect.:3}
\begin{figure*}
	\centering
	\includegraphics[width=\textwidth]{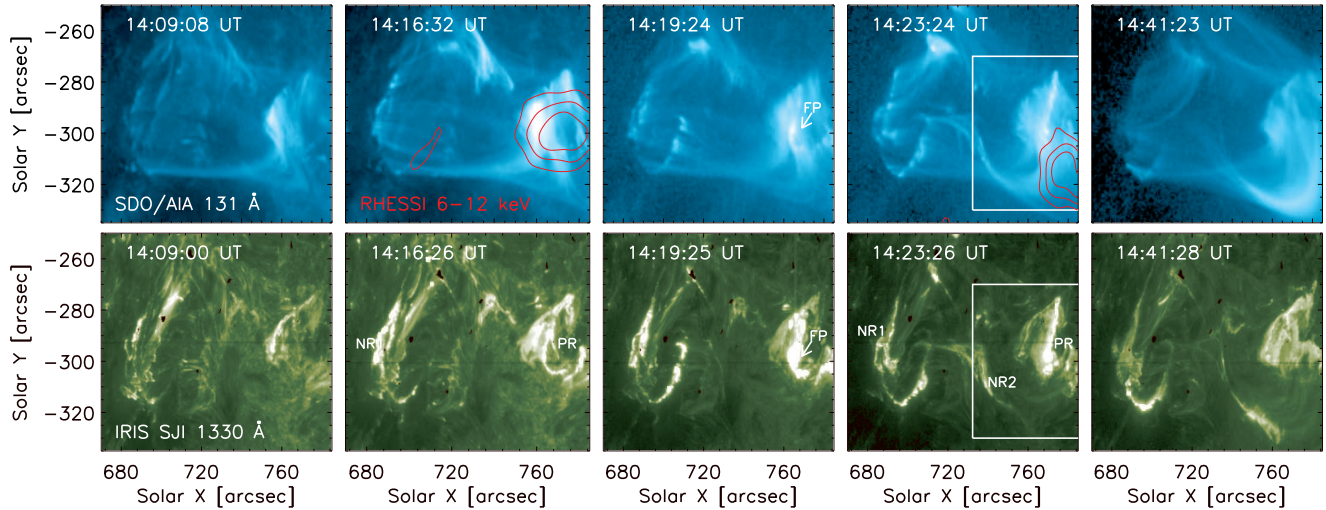} 	
      \caption{Overview of the evolution of the flare loops over time, from 14:09 UT to 14:39 UT. The top panels show the SDO/AIA 131 \AA~images, which are dominated by hot emission from \xxi~during flares. The bottom panels show the closest in time IRIS SJI images in the \cii~filter. Positive polarity (PR) and negative polarity (NR1,NR2) ribbons are labelled in the figure. We focus on studying the flare loops indicated by the white boxes and rooted on the footpoint indicated as \textit{FP}. RHESSI contours of the thermal emission formed in the 6--12 keV channels are overplotted on the second and fourth AIA panels. }
      \label{Fig:overview_131_SJI}
  \end{figure*}
 \begin{figure*}[!ht]
	\centering
	\includegraphics[width=\textwidth, height=40mm]{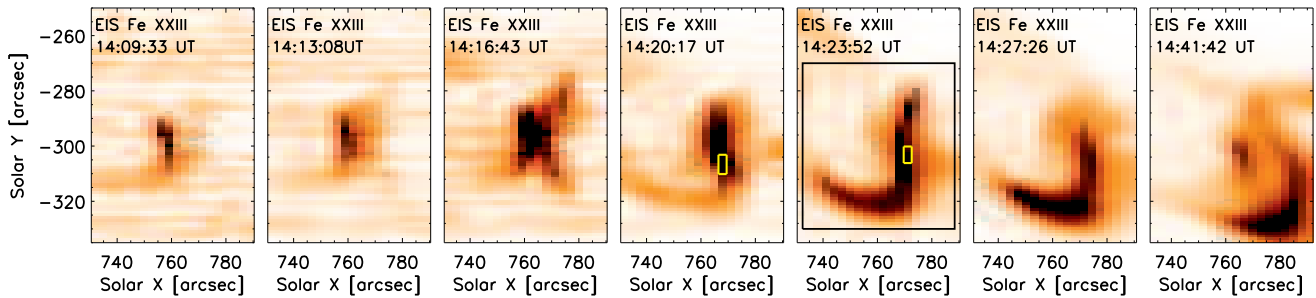} 	
      \caption{Overview of the EIS monochromatic images formed in the \xxiii~263.77~\AA~line. We can observe the evolution of the hot temperature plasma over time. The black box in the fifth panel indicates the flare loops under study. The small yellow boxed areas in the fourth and fifth panels indicate the pixels where we acquired the \xxiii~spectra in Fig.\ref{Fig:XXIII_fits}.}
      \label{Fig:overview_EIS}
  \end{figure*}
The pre-flare state of the AR 12192 is characterized by multiple transient brightenings observed by AIA (see online Movie 1). These brightenings typically last several minutes and have a loop-like morphology in AIA 94~\AA~and 131~\AA, but not in AIA 171~\AA, indicating that their emission originates in \xviii~and higher ionization stage of Fe ions \citep{ODwyer10,Petkaki12,DelZannaWoods13}. The footpoints of these loop-like brightenings are the same as the locations of the flare ribbons during the subsequent X2-class flare. These brightenings indicate ongoing magnetic reconnection along the quasi-separatrix layers \citep{Priest95,Demoulin96,Demoulin97,Titov02, Savcheva2015} involved in the flare.
An overview of the flare evolution, as observed by SDO/AIA in the 131~\AA~band and the IRIS SJI in the 1330~\AA~filter, is shown in Fig.\ref{Fig:overview_131_SJI}.
The IRIS SJI 1330~\AA~images (dominated by \cii~1334/1335\AA~formed at $T \sim 0.02$~MK) show the morphology of the flare ribbons over time. 
The dominant contribution to the AIA 131~\AA~channel during flares comes from \xxi~emission \citep{Petkaki12,DelZanna13,Dudik14b}, which is formed at $\sim$~10~MK.  Therefore, the sequence of AIA images in Fig \ref{Fig:overview_131_SJI} shows the evolution of the high temperature flare plasma and can be directly compared to the EIS \xxiii~monochromatic images in Fig \ref{Fig:overview_EIS}. The latter images show (negative) intensity maps obtained by performing a Gaussian fitting of the \xxiii~spectral line in every pixel of the EIS raster. 

The flare starts with a compact brightening visible in the AIA 131~\AA~image from around 14:09~UT (first panel Fig \ref{Fig:overview_131_SJI}), which corresponds to brightenings all along the positive-polarity ribbon PR in the leading sunspot, visible in the SJI \cii~image. Faint high temperature ($\gtrsim$~10~MK) emission is first observed by IRIS and EIS at the same time, in the very beginning of the rise phase (see EIS image in first panel of Fig. \ref{Fig:overview_EIS}). This is likely to be associated with the flare loops anchored in PR and in the negative-polarity ribbon NR1. A faint continuum emission is observed across the IRIS \oi~ spectral window, which also shows intense and broadened \ci~1354.29~\AA~and \oi~1355.6~\AA~lines at the ribbon PR. The presence of high temperature emission and chromospheric lines enhancement suggests that heating is taking place at low atmospheric heights \citep[c.f.,][]{Graham2013, Simoes2015}. 

After 14:09 UT, intense brightnenings are observed all along the ribbon PR in the IRIS SJI \cii~and AIA 131~\AA~images. The \xxiii~emission observed by EIS also becomes more intense and diffuse along the ribbon, as can be seen in the second panel of Fig. \ref{Fig:overview_EIS}, showing the EIS \xxiii~raster at $\sim$~14:13~UT.

The ribbon PR presents a complex morphology and evolution. It appears to be made up of sub-structures which continuously form, brighten and slip \citep{Dudik14b,Li15} as the flare develops. From about 14:14~UT, an extended branch forms and moves eastward until 14:23~UT. However, the ribbon is visible as a single structure in EIS (see third panel of Fig. \ref{Fig:overview_EIS}) due to the modest spatial resolution of the spectrometer. Between 14:09 and 14:18~UT, we observe a strong thermal emission in the RHESSI 6--­12 keV images, indicating a high temperature of the plasma at the ribbon (second panel of Fig.\ref{Fig:overview_131_SJI}). After 14:18~UT, the thermal HXR source is associated with the hot loops (fourth panel of Fig.\ref{Fig:overview_131_SJI}).

At around 14:19 UT, the ribbon emission becomes very intense around the ribbon position $\sim$~solar $Y$\,=\,-300$\arcsec$ (see the SJI image in the third panel of Fig. \ref{Fig:overview_131_SJI}), which is included within the field of view of IRIS. During the interval $\sim$ 14:18--14:20~UT, we observe strong \xxi~and \xxiii~blue shifts (of the order of 200 km\,s$^{-1}$) originating from a bright footpoint location on PR (indicated as \textit{FP} in Fig. \ref{Fig:overview_131_SJI}). We interpret these upflows as a signature of the plasma evaporating along the flare loops, in accordance with the standard model of flares \citep{Carmichael64,Sturrock68, Hirayama74, KoppPneuman76}. This is consistent with what is shown in the third panel of Fig. \ref{Fig:overview_EIS}, where we can see that the \xxiii~hot emission is at first concentrated at the footpoint location (fourth panel) and then fills a faint flare loop which becomes brighter in the following raster (fifth panel). We note that the high temperature plasma originates predominantly from the footpoint anchored in the positive polarity ribbon, resulting in an asymmetric evaporation. 
  The details of the blue shift emission observed by IRIS and EIS at this location are discussed in Sect. \ref{Sect.:4}. 
  
As the hot plasma evaporates from the ribbon, the flare loops become visible in the AIA 131~\AA~images (white box in Fig.\ref{Fig:overview_131_SJI}), which show the same plasma morphology as the \xxiii~emission observed by EIS (fourth panel of Fig. \ref{Fig:overview_EIS}).  

At 14:23~UT the westward motion of PR ceases, and instead the ribbon exhibits a complex rearrangement with squirming motions (see online Movie 2). At the same time, its long hook \citep{Aulanier12,Janvier13,Janvier14}, extending towards solar $X$\,=\,800$\arcsec$, solar $Y$\,=\,-380$\arcsec$, brightens up (see Fig.\ref{Fig:context}). Additionally, several secondary ribbons \citep[see, e.g.,][]{Chandra09} also appear at this time. These secondary ribbons are connected by flare loops to both PR and NR1. One of these secondary ribbons in the negative polarities is labeled as NR2 in Fig. \ref{Fig:overview_131_SJI}. Its occurrence is at first obscured by bright clumps of downflowing material along loops connecting PR and NR1. Nevertheless, the NR2 can be distinguished visually from 14:23~UT onward. The hot loops under study then expand westward as they connect footpoints progressively moving along PR and NR2, generating an arcade of flare loops. 

We note that due to the complexity of multiple secondary ribbons involved in the flare, there are probably many overlying structures along any given line-of-sight. This seems to be the case also for the bright AIA 131~\AA~loops in the white box in Fig.~\ref{Fig:overview_EIS}. Nevertheless, the flare loops connecting PR and NR2 clearly dominate both the AIA 131~\AA~and EIS signal (see Figs. \ref{Fig:overview_131_SJI}, \ref{Fig:overview_EIS}, as well as the online Movie 2, note the logarithmic scale in intensity).

The flare loops are then observed to cool down in the AIA passbands dominated by progressively lower temperature emission, as can be seen in the online Movie 1. 

In Sect. \ref{Sect.:4}, we focus on the high temperature blue shifts observed by IRIS and EIS at the footpoint of the confined hot loops formed between PR and NR2 included by the boxed areas shown in Fig \ref{Fig:overview_131_SJI} and \ref{Fig:overview_EIS}.

\begin{figure*}[!ht]
	\centering
	\includegraphics[width=0.6\textwidth]{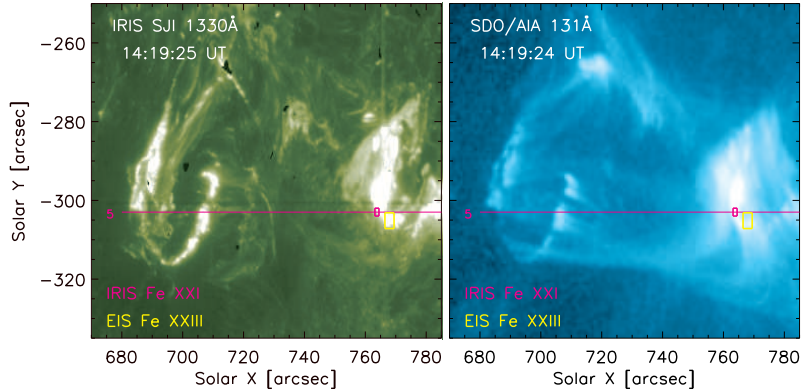} 	
      \caption{IRIS SJI (left) and AIA 131 \AA (right) images showing the footpoint of the flare loops (\textit{FP} in Fig \ref{Fig:overview_131_SJI}) during the impulsive phase. We have focused on the \xxi~blueshifts observed at the slit position no.5, indicated by the horizontal pink line. The pink and yellow boxes represent the positions where respectively the IRIS \xxi~and EIS \xxiii~are observed at around 14:19~UT. See also the online Movie 2 showing the evolution of the SJI and AIA 131 \AA~images over time.}
      \label{Fig:loop_14_19}
  \end{figure*}

\section{Doppler shifts observed by IRIS and EIS}
\label{Sect.:4}
 Blue shifts in high temperature emission lines are simultaneously observed by IRIS (\xxi) and EIS (\xxiii) from around 14:19~UT onward at the position \textit{FP} along the positive polarity (PR) ribbon of the hot flare loops seen in Fig. \ref{Fig:overview_131_SJI}.
 
IRIS was scanning over the footpoint area (\textit{FP} in Fig.\ref{Fig:overview_131_SJI}) mainly at the three slit positions corresponding to the raster exposures No.\,3, 4, 5. The Doppler shifts observed at these positions show similar trends over time. In this study we focus in more detail on the raster exposure No.\,5, where the \xxi~shows the maximum blueshifts and the best fit of the line profile. Its position is indicated by the horizontal pink line overlaid to the AIA~131~\AA~and SJI 1330~\AA~images in Fig. \ref{Fig:loop_14_19}. However, it is important to note that a very strong continuum emission is observed all along the ribbon and especially at the positions observed by the slit exposure No.\,3 and 4. Therefore, we cannot rule out that \xxi~emission is present there but blended with the strong FUV continuum. We conclude that the three IRIS slit positions are likely to observe different footpoints along the PR ribbon. 

\begin{figure}[!htb]
	\centering
	\includegraphics[width=0.5\textwidth, height=120mm]{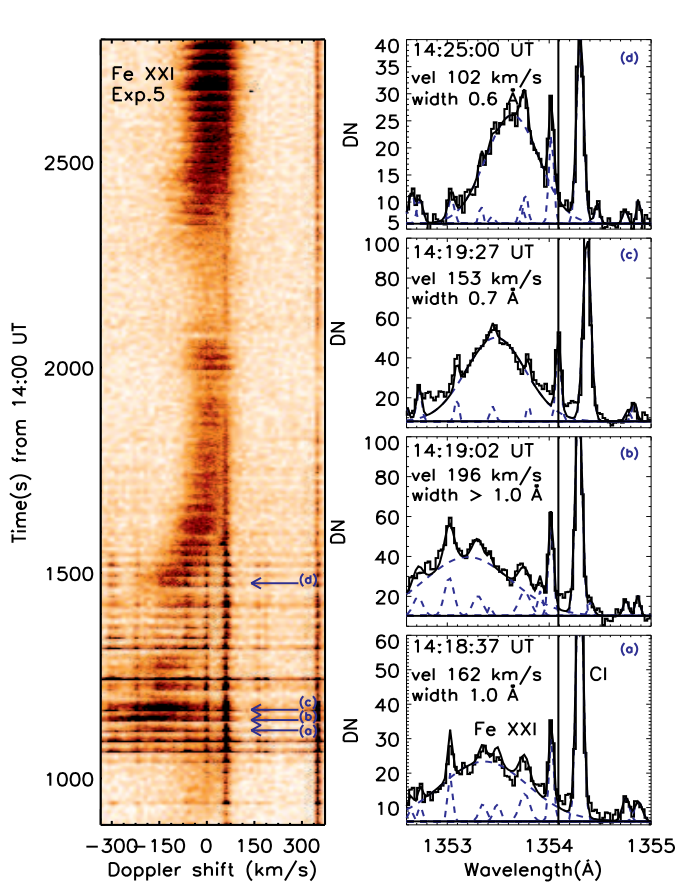} 	
      \caption{Left: IRIS detector images of Fe XXI just above the ribbon as a function of time. The corresponding  velocity values are plotted in Fig. \ref{Fig:Line_profile}. Right: \xxi~spectra at particular times indicated by the blue arrows on the left image. The vertical line in each plot represents the expected rest wavelength position. The fit parameters (centroid velocity and FWHM) are reported for each spectrum. }
      \label{Fig:time_spectra}
  \end{figure}

\begin{figure}[!ht]
	\centering
	\includegraphics[width=0.5\textwidth, height=45mm]{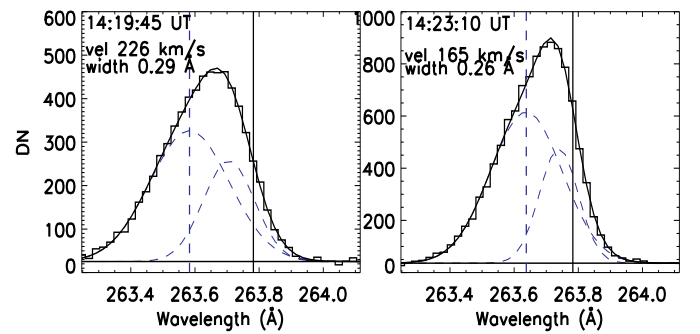} 	
      \caption{EIS \xxiii~profiles in two consecutive rasters during the rise phase of the flare, at about 14:20 UT and 14:23 UT. The vertical line represents the expected rest wavelength position. The fit parameters (velocity and FWHM of the most blue-shifted component) are reported for each spectrum.}
      \label{Fig:XXIII_fits}
  \end{figure}

Fig. \ref{Fig:time_spectra} provides an overview of the \xxi~line profile over time, as observed from the IRIS spectrograph at the raster exposure No.\,5. The left panel represents the velocity-time evolution of the line at the ribbon, obtained by plotting a slice of the FUVS \oi~detector image for each raster from 14:14~UT. Each slice is taken from 2 to 8 pixels in the slit (solar-$Y$) direction just above the position of maximum intensity of the FUV continuum. We note in fact that the \xxi~emission is slightly offset from the FUV continuum location at the ribbon, as already observed by \citet{Young15}. Some of the spectra are shown on the right, corresponding to the time indicated by the horizontal arrow in the velocity-time plot. In Fig. \ref{Fig:XXIII_fits} we show two EIS \xxiii~spectra at the footpoint of the flare loops under study for two consecutive rasters during the peak of the chromospheric evaporation. The locations where we acquired these spectra are indicated by the yellow boxes in Fig. \ref{Fig:overview_EIS}. The parameters of the IRIS and EIS fits (line velocity and FWHM) are given on the top left of each plot in Figs. \ref{Fig:time_spectra} and \ref{Fig:XXIII_fits}. For the EIS \xxiii~spectra, they refer to the most blue-shifted component. In addition, the values of non-thermal width given in the text are calculated as $\sqrt{(4\,\mathrm{ln}2)^{-1}\left(\lambda/c\right)^{-2}\cdot(W^2-W_\mathrm{th}^2-W_\mathrm{I}^2)}$
where $W$ is the line FWHM obtained from the fit, $W_{th}$ is the thermal width of the line calculated by using the atomic parameters from CHIANTI v7.1, $W_\mathrm{I}$ is the instrumental FWHM, $\lambda$ is the rest wavelength and c is the speed of light.

Fig. \ref{Fig:Line_profile} summarizes the results of the Doppler shifts observed by IRIS and EIS during the impulsive phase of the flare. In particular, the line position of \xxi~(orange), \xxiii~(blue, triangle) and \siiv~(green) are plotted as a function of time. The \xxiii~velocity values refer to the fit results of the most blue-shifted component in the EIS spectra shown in Fig. \ref{Fig:XXIII_fits}. The \xxi~and \siiv~centroid positions have been obtained by fitting the spectra averaged over each detector slice in Fig. \ref{Fig:time_spectra} over time. 

  We will now discuss in details the high temperature blue shifts as observed simultaneously (within $\sim$~20\,s) and co-spatially (within the alignment uncertainty, 3--5$\arcsec$) by both spectrometers.

\subsection{Early phase, 14:14--14:17~UT}
At 14:14:47 UT (corresponding to the IRIS raster No.\,24) we observe a sudden intensity enhancement of the chromospheric \oi~and \ci~lines at the ribbon PR. A faint FUV continuum can also be observed all along the spectral window. This is likely to suggest that a heat deposition is taking place in the chromosphere. No significant \xxi~emission is observed at this time at the positions observed by the IRIS slit. From 14:16:04 UT (IRIS raster no.27) we observe very faint \xxi~emission ($\sim$ 10 DN) above the positive ribbon (PR), which can be seen in the bottom part of the velocity-time plot in Fig \ref{Fig:time_spectra} (Time $<$ 1000 s). The line position at that time is blue-shifted by around 30--45 km/s and the line width is around 0.8--0.9 \AA~(corresponding to a non-thermal width of $\sim$~80--100~km\,s$^{-1}$). This high temperature emission is likely to originate from a separate bundle of flare loops forming between PR and NR1 (see second panel of Fig \ref{Fig:overview_131_SJI} and online Movie 1). Therefore, we are not considering this early \xxi~emission to have originated from the evaporation of hot plasma from the footpoint \textit{FP}.

\subsection{Peak of the impulsive phase, 14:17--14:20~UT}
The FUV continuum emission at the ribbon then suddenly becomes very intense from 14:17:20--14:17:46~UT, corresponding to the IRIS raster No.\,30--31 onward. Due to the very strong continuum emission from the ribbon, we are not able to detect a good \xxi~line profile for around 50s after the first ribbon intensity enhancement takes place.  Therefore,  we cannot rule out that very faint blue-shifted \xxi~emission was present at this time, but covered by the enhanced continuum intensity from the ribbon. 

 At 14:18:37 UT (raster No.\,33), large blue-shifted \xxi~emission from the flare footpoint is observed by the IRIS slit, indicated in the velocity-plot in Fig. \ref{Fig:time_spectra} by the first blue arrow (a). The corresponding spectrum is shown on the right. The line shows a very broad profile, with a width of 1~\AA~(corresponding to a non thermal velocity of $\sim$ 130 km\,s$^{-1}$) and completely blue-shifted by $\sim$ 110 km\,s$^{-1}$. In the following raster at 14:19:02 UT, the line profile presents a larger blueshift, which seems to partially expand outside the wavelength range of the spectral window. Its spectrum (b) is shown in Fig \ref{Fig:time_spectra} (second panel from the bottom). It is not possible to fit the entire Gaussian profile and therefore the blueshift and line width values (196 km\,s$^{-1}$, 1~\AA) shown in the figure might represent a lower estimate. 
 
 It is interesting to note that the line profile is less blue-shifted initially ($\sim$~160~km~s$^{-1}$, (a)) before reaching the largest Doppler shift value ($\sim$~200~km~s$^{-1}$, (b)). Similarly, a high cadence EIS flare study ($\sim$ 11 s) analyzed by \citet{Brosius13} reported \xxiii~blue-shifted profiles which became even more blue-shifted over 2 exposures before gradually decreasing to zero in about 12 exposures. To our knowledge, this is the first time that a similar feature was observed with the improved resolution of the IRIS spectrometer. However, the observation of different Doppler shifts along the line of sight might also be the result of a change in the inclination of the local magnetic field at the position of the IRIS slit. More observational data would confirm if this trend was related to the real evolution of the evaporation velocities or caused by a geometric effects. Unfortunately, as we pointed out above, we cannot reliably measure the \xxi~line blueshift before 14:18:37~UT. 
 
At around 14:19:27 UT, the \xxi~line at the footpoint \textit{FP} is completely blue-shifted by $\sim$ 150 km\,s$^{-1}$ with a line width of $\sim$ 0.70 \AA (corresponding to a non thermal velocity of around 60 km\,s$^{-1}$) as shown in in the third spectrum (c) in Fig \ref{Fig:time_spectra}. The EIS slit was scanning the same area during the raster exposure no.6, at $\sim$ 14:19:45~UT. 
The \xxiii~line has an asymmetric profile with an enhanced blue wing indicative of plasma upflows of $\sim$ 226~km\,s$^{-1}$ and a second less blue-shifted profile at $\sim$ 90 km\,s$^{-1}$. The most blue-shifted component shows a very broad line profile with a width of 0.29~\AA~(non thermal width of $\sim$ 180 km\,s$^{-1}$). The spectrum is shown in the left panel of Fig.\ref{Fig:XXIII_fits}. This represents the closest observation in time ($\sim$~20~s) and space (within the 3--5$\arcsec$ alignment uncertainty) where both spectrometers were scanning over the flare footpoint \textit{FP} during the peak of the chromospheric evaporation. The locations of the observed IRIS \xxi~and EIS \xxiii~blueshifts at $\sim$ 14:19~UT are indicated in Fig.\ref{Fig:loop_14_19} by the pink and yellow boxes respectively overlaid to the closest AIA 131~\AA~and SJI~\cii~images.

The interpretation of the \xxiii~multi components profile is not straightforward. It is important to note that, due to the longer raster cadence, EIS missed the first minute of the early chromospheric evaporation flows observed by IRIS from $\sim$~14:18:30~UT. In addition, we note that the bundle of loop structures formed between PR and NR1 are very close (few arcseconds) to the footpoint location. Therefore, it is likely that the EIS slit is observing a superposition of hot plasma flows from either different sub-resolution loop strands or distinct locations along the line of sight, as already suggested in some EIS studies \citep[e.g.,][]{Milligan09}. 

\begin{figure}[!htl]
	\centering
	\includegraphics[width=0.35\textwidth]{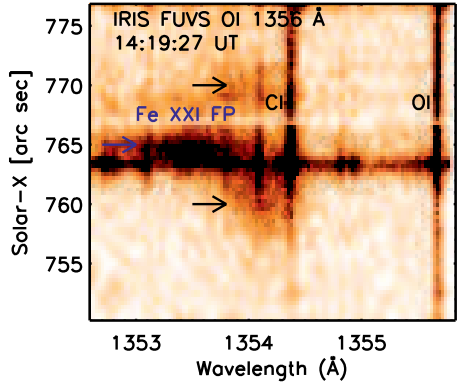} 	
      \caption{IRIS Spectrograph FUVS detector image (solar-$X$ vs wavelength) of the \oi~1356 spectral window. A blue-shifted \xxi~1354.08~\AA~line profile is particularly visible at around solar-$X$ = 765$\arcsec$, as indicated by the blue arrow. Faint \xxi~emission close to rest is also visible at other locations along the slit, as indicated by the black arrows. }
      \label{Fig:detector}
  \end{figure}

In contrast, the \xxi~line profiles observed by IRIS are symmetric and completely blue-shifted during all the observation. The IRIS \xxi~observations are thus consistent with what is predicted by 1D hydrodynamics models, that a single blue-shifted component should be observed during the impulsive phase of the flare \citep{Emslie87}. However, we point out that in the study of a small C-class flare, \cite{Polito15} observed asymmetric \xxi~profiles after the evaporation first occurred. They interpreted the observed blue wing as due to the newly evaporating plasma superimposing on the \xxi~which had already filled the loop. This would suggest that if the size of the footpoint area is small enough compared to the IRIS spatial resolution, we might not be able to distinguish plasma flows originating from different loop strands. For instance, Fig \ref{Fig:detector} shows an image of the FUVS \oi~detector image at 14:19:27~UT, where we can observe distinct \xxi~emission along the solar-$X$ slit direction, as indicated by the blue (blue-shifted plasma) and black (plasma close to the rest wavelength) horizontal arrows. These sources are separated by $\sim$ 3$\arcsec$ or less. Hence, they might not be resolved by the EIS spectrometer, whose point spread function is around 3$\arcsec$. In contrast, it appears that the IRIS spatial resolution is high enough to resolve different locations when we are observing enough large size flares, as in the present study.

Another possible explanation for the multi-component profiles observed by EIS might be due to fact that the exposure time is long compared to the evaporation timescale and that we are averaging along plasma upflows at different times. We note that completely blue-shifted IRIS \xxi~profiles were also reported by \citet{Young15},\citet{Tian15} and \citet{Graham15} during the whole evolution of two X-class flares with an exposure time of $\sim$ 9 s, which is the same as the EIS exposure time used in the present study. We would thus suggest that asymmetric high temperature line profiles observed by EIS and not by IRIS are are likely to be due to the lower spatial rather than temporal resolution of EIS. 
\begin{figure}[!h]
	\centering
	\includegraphics[width=0.5\textwidth]{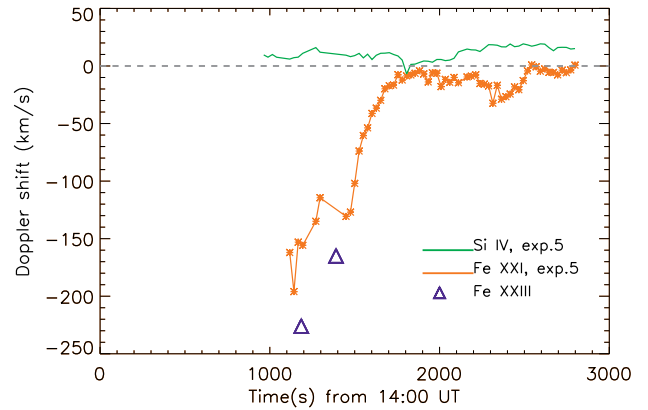} 	
      \caption{Doppler shift as a function of time of IRIS~1354.08~\AA,~\siiv~1402.77~\AA~and EIS \xxiii~263.77~\AA. In Sect. \ref{Sect.:6.3}, we compare the values of velocity derived from observations with the simulated results of Fig.\ref{Fig:shifts}.
}
      \label{Fig:Line_profile}
  \end{figure}  
  
\subsection{Peak and gradual phase, 14:22~UT onward} 
From about 14:22~UT--14:23~UT, bright hot flare loops are clearly visible in the AIA 131~\AA~and EIS \xxiii~images, as shown in Figs.\ref{Fig:overview_131_SJI} and \ref{Fig:overview_EIS}. They are formed between the footpoint location \textit{FP} and the secondary negative ribbon NR2, shown in the HMI image in Fig. \ref{Fig:context}. As the evaporation proceeds, we observe that the ribbon PR rapidly shows intense compact brightenings all along its length. The \xxi~spectra at the footpoint position (\textit{FP}) are progressively less blue-shifted, as shown in the velocity-time plot of Fig \ref{Fig:time_spectra}, indicating that the evaporation speed is slowly decreasing. The EIS slit crossed the same footpoint region at around 14:23:10~UT. The corresponding spectra is shown in the second panel of Fig \ref{Fig:XXIII_fits}. The \xxiii~line profile is asymmetric with a dominant more blue-shifted component at $\sim$ 165 km\,s$^{-1}$ and with a line width of $\sim$ 0.26~\AA.

 At the same time ($\sim$~14:23~UT), the PR ribbon shows intense brightnenings all along its length at solar-$Y$ = -290/-280$\arcsec$, as can be seen better in the fourth panel of Fig.\ref{Fig:overview_131_SJI} and in the online Movie 2. We note that at this location the  \xxiii~line is completely blue-shifted up to $\sim$ 180 km\,s$^{-1}$. It is not clear if these \xxiii~upflows might indicate a secondary footpoint(s) source for the flare loops under study. However, its location is much higher than the footpoint \textit{FP} indicated in Fig \ref{Fig:overview_131_SJI}, where we compare \xxi~and \xxiii~upflows.
 
Finally, it is interesting to note that at the top of the faint hot loops visible at around 14:20~UT (third panel of Fig \ref{Fig:overview_EIS}) the \xxiii~line appears to be significantly broad (0.23~\AA) and slightly blue-shifted ($\sim$ 36.4 km\,s$^{-1}$). This could indicate plasma turbulence during this early impulsive phase, when the hot loops are being filled by evaporating plasma from the footpoints. The line width at the loop tops then progressively decreased going towards the peak of the flare, when the emission from the hot loops becomes more intense. For instance, at 14:39~UT the width of the \xxiii~line profile at the top of the flare loops is almost thermal ($\sim$ 0.13~\AA) and at rest ($\sim$~5.7~km\,s$^{-1}$). 

Finally, we emphasize that the flare under study is not located at the center of the disk, and thus the effect of the line of sight might reduce the size of the Doppler shifts observed. For instance, \citet{Tian15} and \citet{Graham15} observed \xxi~blueshifts of up to approximatively 300~km~s$^{-1}$ during the impulsive phase of the 10 September 2014 X-class flare, which would be consistent with the values found in the present study, considering an inclination of $\sim$ 25--30 degrees along the line of sight.
 \subsection{The \siiv~Doppler shifts with IRIS}
 At the same location along the positive polarity ribbon (IRIS raster slit position No.\,5 ), we also studied the evolution of the \siiv~1402.77~\AA~line observed by IRIS. The line appears to be significantly broader during the impulsive phase of the flare, but does not show significant red wing asymmetries, contrary to what has been reported by \citet{Tian15} in a recent flare observation with IRIS. Before the first flare peak at around 14:25~UT, the line position is redshifted by $\sim$~8~km~s$^{-1}$, which exceeds by about 3~km~s$^{-1}$ the $\sim$~5~km~s$^{-1}$ Doppler shift observed in the quiet Sun \citep{Peter99}. However, the wavelength calibration of the FUVL wavelength array is based on the measurement of the \siv~line centroid, which gives an error of 2 km\,s$^{-1}$ or more, depending on the signal/noise of the line. We observe a significant redshift ($\sim$~16~km~s$^{-1}$) only at around 14:18~UT, at the peak of the FUV continuum intensity at the flare ribbon.
 
During the gradual phase we then observe a \siiv~redshift at the PR ribbon of around 17~km~s$^{-1}$, which is likely to be due to the condensation and downflow of the cool and dense plasma along the flare loop.

\section{Plasma diagnostics from EIS/IRIS/AIA}
\label{Sect.:5}
Spectroscopy and imaging in the EUV/UV emission provide a powerful tool to diagnose the plasma physical parameters during flares. In this section, we shall investigate the values of density, temperature and emission measure of the flare ribbons and loops by combining the sets of observation from IRIS, EIS, XRT and AIA. In Sect. \ref{Sect.:6.3}, we will then compare the observed diagnostics with the results predicted by hydrodynamics simulation with the HYRAD code.

\subsection{Electron density diagnostics with IRIS }
\label{Sect.:5.1}
\begin{figure*}[!ht]
	\centering
	\includegraphics[width=\textwidth]{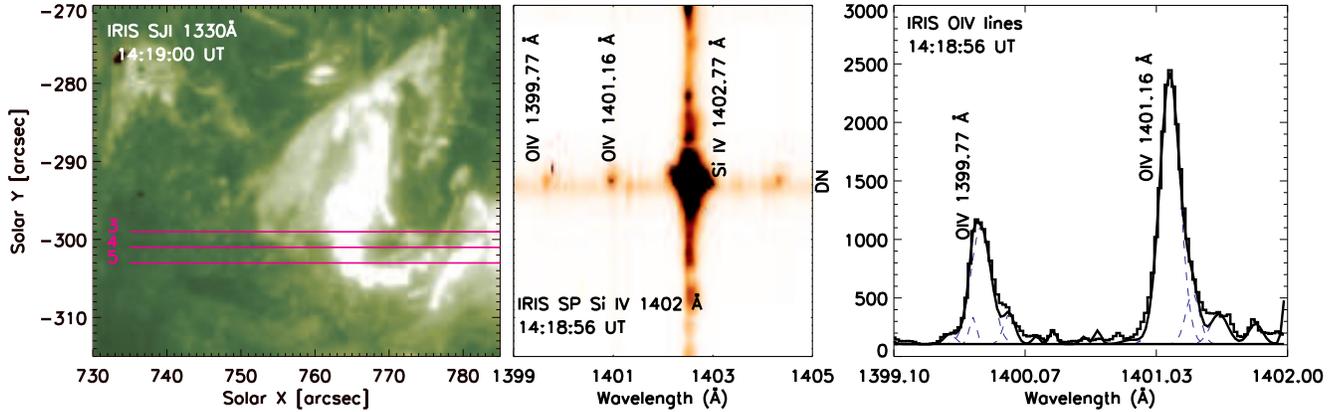} 	
      \caption{Left: IRIS SJI 1300 image with the positions of the IRIS slit overplotted.  Middle: Corresponding detector image of the Si IV spectral window with the OIV lines. Right: Sample spectrum of the OIV lines at the ribbon location corresponding to the slit position No.\,3.}
      \label{Fig:density_OIV}
  \end{figure*}

The ratio of the IRIS \oiv~1399.77~\AA~and 1401.16~\AA~lines is sensitive to the electron number density of the plasma from which they are emitted and therefore provides an important density diagnostic for the transition region plasma during flares. 
In solar active regions, the \oiv~emission (formed at log($T$[K])~$\sim$~5.2) is often very low compared to other lines formed at similar temperatures, such as the \siiv~1402.77~\AA~(log($T$[K])~$\sim$~4.8 ). However, during the impulsive phase of flares, the \oiv~emission is usually enhanced at the ribbon and can therefore be reliably measured. Fig \ref{Fig:density_OIV} shows an example of the IRIS \oiv~lines during the impulsive phase of the 27 October 2014 flare at one ribbon location. The first panel on the left shows the IRIS SJI \cii~1330~\AA~image with overplotted the positions of the IRIS slit exposures No.\,3, 4, and 5, which observe the flare ribbon. A sample of the \siiv~detector image (with the \oiv~lines) for the exposure No\,.3 is shown in the second panel. The corresponding \oiv~lines spectrum at a compact footpoint location (clearly visible in the detector image) is plotted in the third panel. 
The spectral lines have been fitted with Gaussian profiles. The 1399.77/1401.16\AA~ratio has been obtained after calibrating in physical units (erg s$^{-1}$ sr$^{-1}$ cm$^{-2}$ \AA$^{-1}$) the line intensities given by the Gaussian fit.
  
\begin{figure}[!ht]
	\centering
	\includegraphics[width=0.5\textwidth, height=65mm]{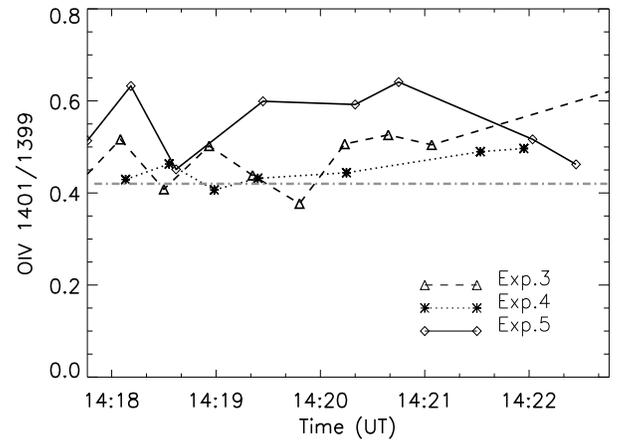} 	
      \caption{\oiv~1399/1401~\AA~ratio at three IRIS slit locations along the ribbon during the impulsive phase of the flare, indicated in Fig.\ref{Fig:OIV_spectrum}. The CHIANTI  v7.1 high density limit is indicated by the horizontal dot-dash line.}
      \label{Fig:ratio_OIV}
  \end{figure}

 The radiometric calibration of the IRIS throughput is detailed in the IRIS technical note 26~\footnotemark[2]. In particular, we used the updated version of the \textit{solarsoft} routine \emph{iris\_get\_response.pro} which takes into account the instrumental degradation since launch. Fig \ref{Fig:ratio_OIV} shows the IRIS \oiv~ 1399.77/1401.16\AA~ratio at the three slit positions along the ribbon (no. 3, 4, 5 indicated in the SJI image in fig. \ref{Fig:density_OIV}) from $\sim$ 14:18~UT to 14:22~UT. The horizontal dotted line indicates a ratio of 0.42, which corresponds to the high density limit of $10^{12}$ cm$^{-3}$ calculated in CHIANTI v7.1, assuming ionization equilibrium. 
We note that the measured \oiv~ratio is above the high density limit almost everywhere, which would indicate an electron density close or higher than $10^{12}$ cm$^{-3}$. 
However, there are at least two issues that we need to take into account when diagnosing the electron density from the IRIS \oiv~1399.77~\AA~and 1401.16~\AA~ratio. First of all, the density sensitivity of the line ratio  depends significantly on the assumption of equilibrium conditions of the plasma. For instance, a ratio higher than 0.42 could also indicate a departure from an electron Maxwellian distribution in the observed plasma \citep{Dudik14a}. Other observables would be needed to constraint the non-equilibrium plasma conditions and disentangle the two effects.

\begin{figure}[!ht]
	\centering
	\includegraphics[width=0.5\textwidth]{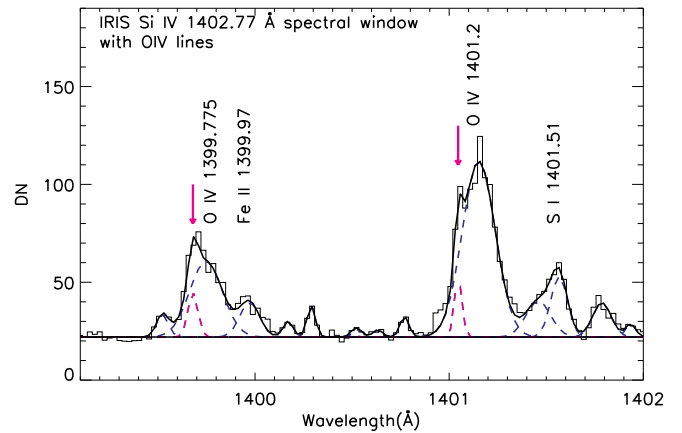} 	
      \caption{An example of a spectrum of the \oiv~lines observed by IRIS showing the blends with cool narrow emission lines in the same spectral range. The unidentified lines are indicated by the pink arrows.}
      \label{Fig:OIV_spectrum}
  \end{figure}

Another important issue when measuring line intensities is represented by possible blendings of the line profiles. In this study, we noted that the emission from unidentified cool emission lines in the IRIS FUVL spectral range is enhanced during the impulsive phase of the flare. These lines blend with the \oiv~lines and  can cause an additional uncertainty in the density diagnostics. An example is reported in Fig \ref{Fig:OIV_spectrum}, which shows a spectrum of the \oiv~lines blended with cool unidentified emission lines, indicated by the pink arrows. 
The main identified blending affecting the \oiv~1399.77~\AA~line is the \feii~1399.62~\AA. In the present study, we observed that an unidentified (probably chromospheric) narrow emission line blends the blue wing of the \oiv~1399.77~\AA~at around 1399.70~\AA. The emission from this line is only enhanced at the ribbon location during the impulsive phase of the flare.\\ The \oiv~1401.77~\AA~is usually considered free of blends. However, we identified a narrow spectral line at 1401.05~\AA~blending the \oiv~1401.77~\AA~line, as shown in Fig \ref{Fig:OIV_spectrum}. Finally, the \si~1401.51~\AA~line partly blends with a spectral line at $\sim$~1401.40~\AA~, which could correspond to the unidentified line at the same wavelength reported by \citet{Sandlin86}. 

Thanks to the high instrumental capabilities of the IRIS spectrometers, we are now able to study the flare spectra in the wavelength range $\sim$~1399--1406~\AA~with unprecedented resolution and temporal cadence. To our knowledge, this is the first time that such unidentified blends with the \oiv~lines have been observed. We would like to point out that it is extremely important to take into account and report all these blends when performing plasma diagnostics using the \oiv~lines observed by IRIS.
\footnotetext[3]{http://iris.lmsal.com/documents.html}

  \begin{figure*}[!ht]
	\centering
	\includegraphics[width=\textwidth]{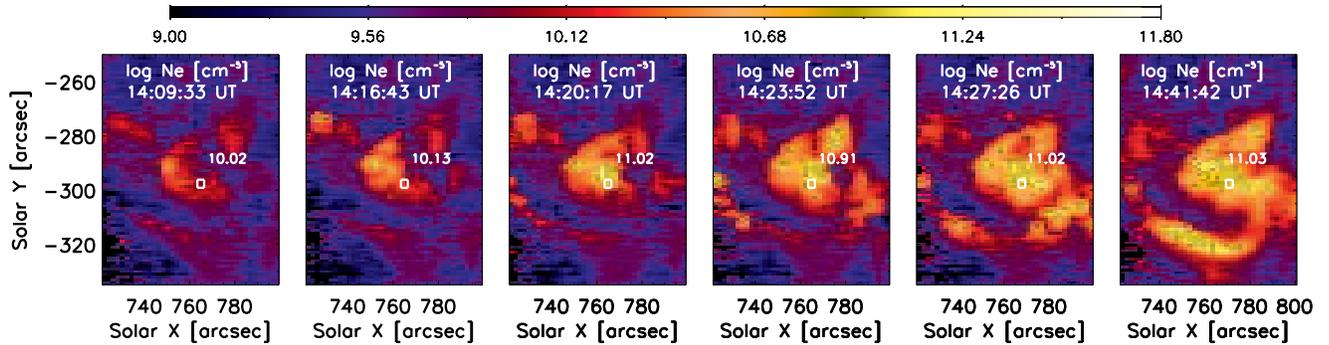} 	
      \caption{Electron densities evolution obtained from the \xiv~264.79/274.20~\AA~line ratio for the 27 October 2014 flare. The colour maps show the log($N_\mathrm{e}$[cm$^{-3}$]). The values of log($N_\mathrm{e}$[cm$^{-3}$]) averaged over the boxed areas at the flare ribbon are reported for each panel.}
      \label{Fig:density_FeXIV}
  \end{figure*}

\subsection{Electron density diagnostics with EIS }
\label{Sect.:5.2}
We obtained electron densities from the ratio of the EIS \xiv~264.79 and 274.20~\AA~lines, using the \xiv~atomic calculations included in the CHIANTI v7.1 database. The \xiv~line at 264.79~\AA~is free of significant blends in active region spectra \citep{DelZanna06}. The \xiv~274.20~\AA~is blended with a \sivii~line at 274.175~\AA. One way to remove this blending is to estimate the \sivii~274.175~\AA~contribution from the intensity of unblended \sivii~lines observed by EIS. Unfortunately, other \sivii~lines were not included in this study. However, during flares the \sivii~contribution to the \xiv~274.203~\AA~is usually estimated to be small, of the order of 4 $\%$ \citep{DelZanna06,Brosius13}. Therefore, it is important to keep in mind that we are providing a lower value of the electron density due to such unknown (but small) \sivii~contribution to the \xiv~274.20~\AA~line.

Fig. \ref{Fig:density_FeXIV} shows the electron density evolution during the impulsive phase. Each panel represents the density map (solar $X$ vs solar $Y$) obtained from the \xiv~264.79/274.20\AA~ratio for a single EIS raster. The colour bar shows the log($N_\mathrm{e}$[cm$^{-3}$]) and the overlaid white box indicates a small area along the PR ribbon close to the footpoint position \textit{FP} where the blueshifts originate. The box does not include the few pixels where we observe the maximum of the \xxiii~intensity because most of the emission lines included in the EIS study are saturated there. We note that the position of the footpoint emission moves over time during the evolution of the flare. The density values averaged over the small box are indicated on each EIS raster.

At around 14:09~UT, in the very beginning of the impulsive phase, we observe a density of $\sim$ $1 \cdot 10^{10}$~cm$^{-3}$ at the flare ribbon, which gradually increase going towards the peak of the flare. In particular, at approximately 14:20~UT (when we observe the largest blueshifts in the high temperature lines) the density of the 2~MK plasma  is around $10^{11}$~cm$^{-3}$. It then remains constant at the value of $\sim$ $10^{11}$ cm$^{-3}$. Later on at around 14:40~UT, close to the second peak of the flare, it is interesting to observe the dense plasma emission from the previously hot loop structures which have now cooled down to 2~MK.

\begin{figure}[!ht]
	\centering
	\includegraphics[width=0.4\textwidth, height=200mm]{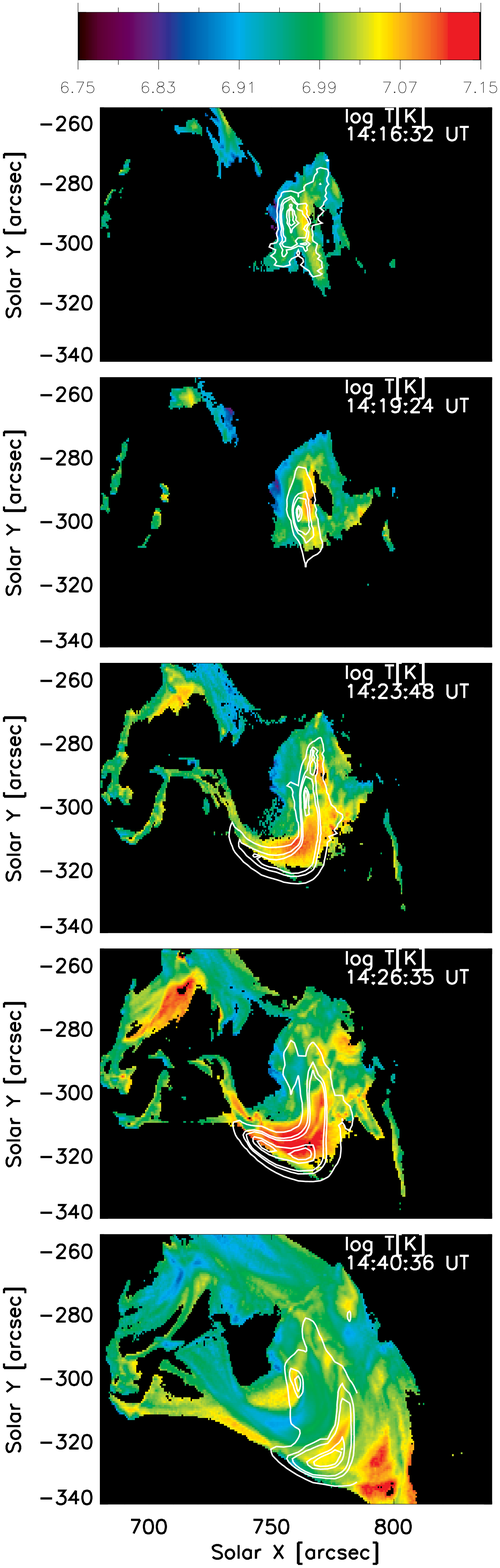} 	
      \caption{Temperature estimation from the ratio of AIA 131~\AA~and~94~\AA channels on the 27 October 2014 flare. The colormaps show the log($T$[K]). Overplot are the EIS \xxiii~intensity contours.}
      \label{Fig:temp_maps}
  \end{figure} 
   
\begin{figure*}[!ht]
\centering
\includegraphics[width=0.8\textwidth]{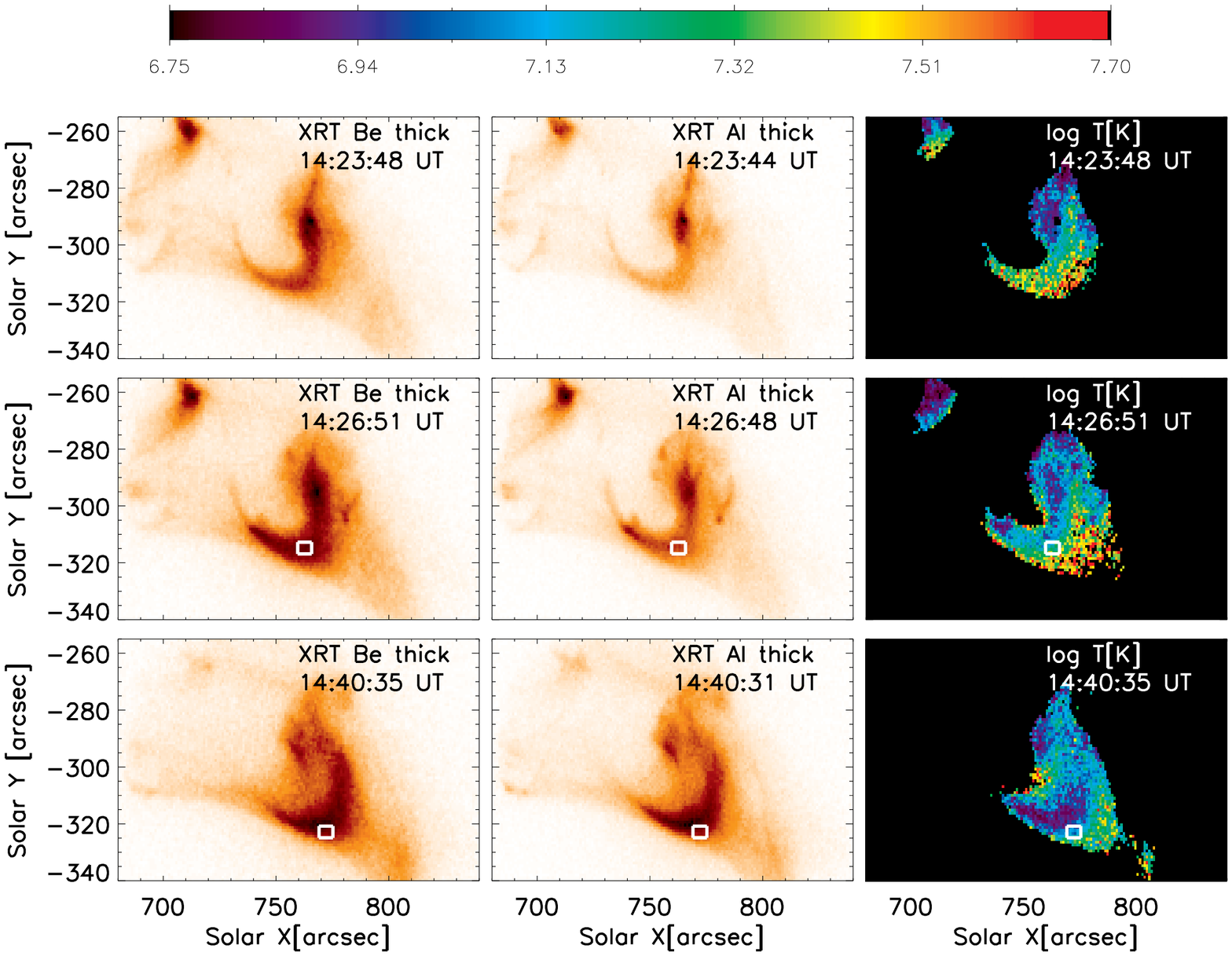}	
      
\caption{Temperature measurements (log($T$[K]),right columns) from the ratio of XRT Be\_thick (left columns) and Al\_thick (middle columns) of the flare loops at three specific times. The temperature in the boxed areas at 14:27~UT and 14:40~UT is $\sim$ 18.7~MK and 13.5~MK respectively, in agreement with EM measurements in Sect. \ref{Sect.:5.4}.}
 \label{Fig:XRT_maps}
  \end{figure*}
  
\subsection{Temperature diagnostics with SDO/AIA and Hinode/XRT}
\label{Sect.:5.3}
During flares, the ratio of the 94 \AA~and 131 \AA~AIA channels can reliably be used to provide temperature diagnostics. For the hot flaring plasma, these channels are in fact dominated by \xviii~(formed at $\sim$ 7 MK) and \xxi~(formed at $\sim$ 10 MK) respectively \citep[see, e.g.,][]{Petkaki12,DelZanna13,Dudik14b}.
Fig. \ref{Fig:temp_maps} shows the evolution of the isothermal temperature obtained by comparing the theoretical ratio of these two bands with the observed intensities. The theoretical AIA 94 \AA~and 131 \AA~countrates were obtained by convolving the AIA effective areas (from the solarsoft \textit{aia\_get\_response.pro} routine) with the CHIANTI synthetic spectra, as detailed in the appendix of \citet{DelZanna11b}.

The colour maps represent the log($T$[K]) in the range 6.75 to 7.2 over which the ratio of the AIA bands \AA~131/94 is sensitive to the high temperature emission. Overplotted are the contours of the EIS \xxiii~intensity (20, 40, 50, 80, 90 $\%$ of the maximum intensity) showing the morphology of the hot flare loops. The contribution of the low temperature emission to the 94 \AA~passbands has been removed by using a combination of the 211 \AA~and 171 \AA~filters, which are sensitive to cool plasma, as detailed in \citet{DelZanna13}. For the 131 \AA~channel, we subtracted a background image which we took to be at the time frame which has the lowest emission in this channel for this event.

The first two panels of Fig.\ref{Fig:temp_maps} ($\sim$~14:16~and 14:19~UT) show that the high temperature emission (log($T$[K])~$\sim$ 7--7.20) is concentrated at the flare PR ribbon during the early impulsive phase of the flare, in agreement with the recent observations by \citet{Simoes2015}. Within the alignment uncertainty, this is the location where the hot \xxiii~emission originates, as observed by the EIS slit. Going towards the first peak of the flare (third and fourth images in Fig.\ref{Fig:temp_maps}), we can see the flare loops which have been filled by the hot plasma evaporating from the ribbon. They overlap with the \xxiii~263.77~\AA~loop structures visible by EIS. It is interesting to note that hot plasma is now also visible at the eastern footpoints along the secondary ribbon NR2. The $\sim$ 11 MK flare loops then progressively move rightward, as the post-flare loops arcade develops, as shown in the last panel taken at around 14:40 UT. 

Observations in multiple XRT bands can also provide temperature measurements in flares \citep{Narukage2011, Narukage2014,ODwyer2014}. We compare the temperature diagnostic obtained with AIA with that derived by using the ratio of the XRT Be\_thick and Al\_thick images (Fig. \ref{Fig:XRT_maps}). We calculated the CHIANTI isothermal spectra using the same set of abundances \citep{Asplund09}, ionization equilibrium calculations \citep{Dere2009} and density (1$\cdot$ 10$^{11}$ cm$^{-3}$) that we used to produce the AIA 131~\AA~and 94~\AA~count rates. We then convolved these spectra with the effective area of each XRT channel,  provided by the solarsoft \textit{make\_xrt\_wave\_resp.pro} routine, which takes into account the contamination of the instrument filters over time.

Fig. \ref{Fig:XRT_maps} shows the temperature maps from XRT at three time intervals during the flare, that we can directly compare with the AIA temperature measurements in Fig. \ref{Fig:temp_maps}. 
The temperature of the flare loops calculated by XRT are higher (log($T$[K])~$\sim$ 7.3--7.4) at the top of the flare loops than the one derived from AIA observations (note the different colour-scales in Figs. \ref{Fig:temp_maps} and \ref{Fig:XRT_maps}).  We emphasize that the ratio of the AIA channels reaches the high temperature limit of log($T$[K])~=~7.2, indicating that even higher temperature plasma might be present at the flare looptop, as is expected from both the solar flare model in 2D \citep{Carmichael64,Sturrock68, Hirayama74, KoppPneuman76} and 3D \citep{Aulanier12}.
 At around 14:27~UT and 14:40~UT (second and third rows in Fig. \ref{Fig:XRT_maps}), XRT observes temperatures of $\sim$~18.7~MK and 13.5~MK in the small boxed area at the flare loop top, respectively. These values suggest a decrease of $\sim$ 4-5 MK of the loop top temperature between the two peaks of the flare ($\sim$ 14:30~UT and 14:40~UT in the soft X-rays curves, Fig. \ref{Fig:goes}). The XRT temperatures are also consistent (within $\sim$~15\%) with the results of the EM loci measurements in Sect. \ref{Sect.:5.4} and with the hydrodynamics simulations in Sect. \ref{Sect.:6}.  
\begin{figure}[!ht]

\includegraphics[width=0.5\textwidth]{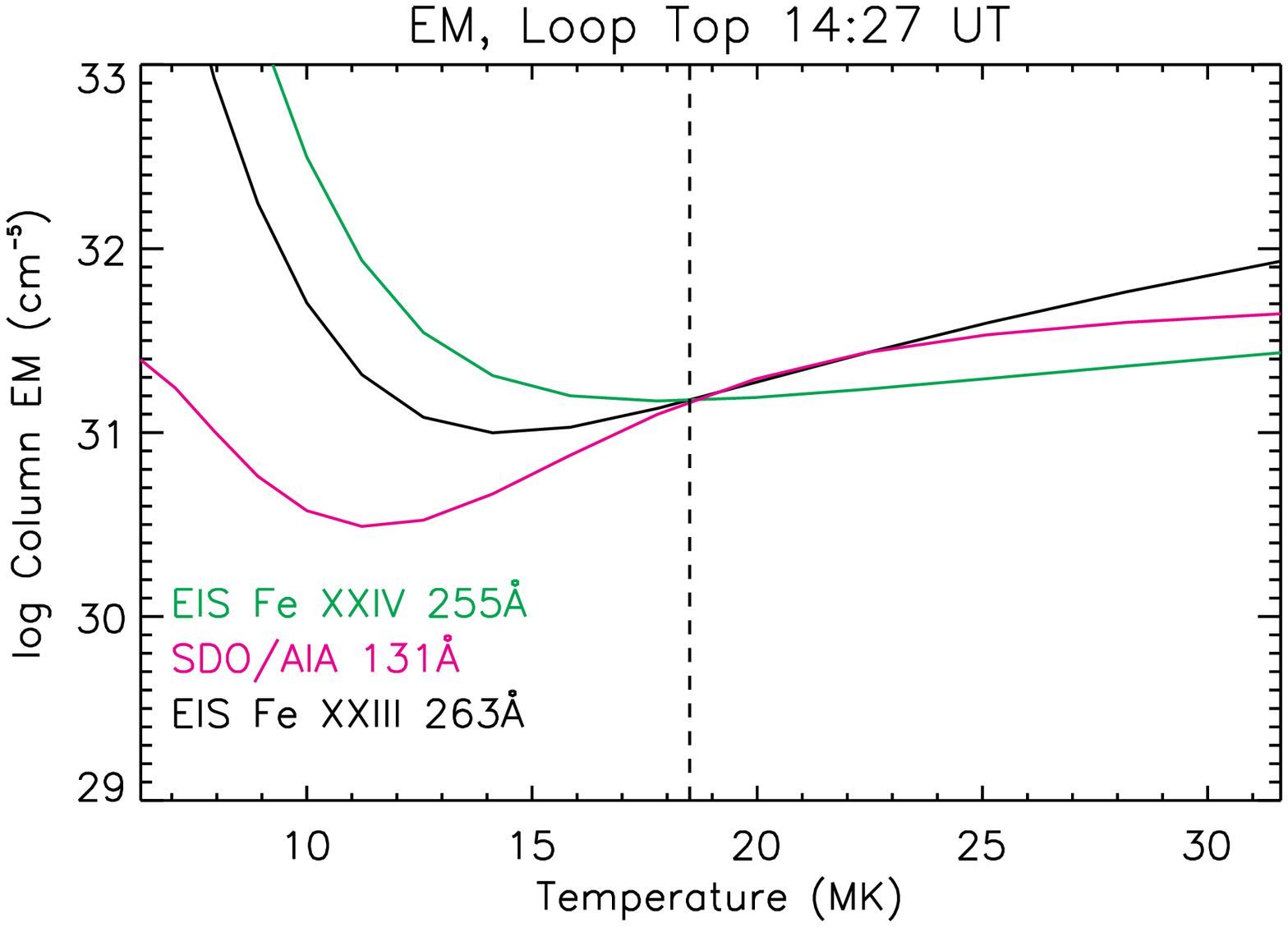}
\includegraphics[width=0.5\textwidth]{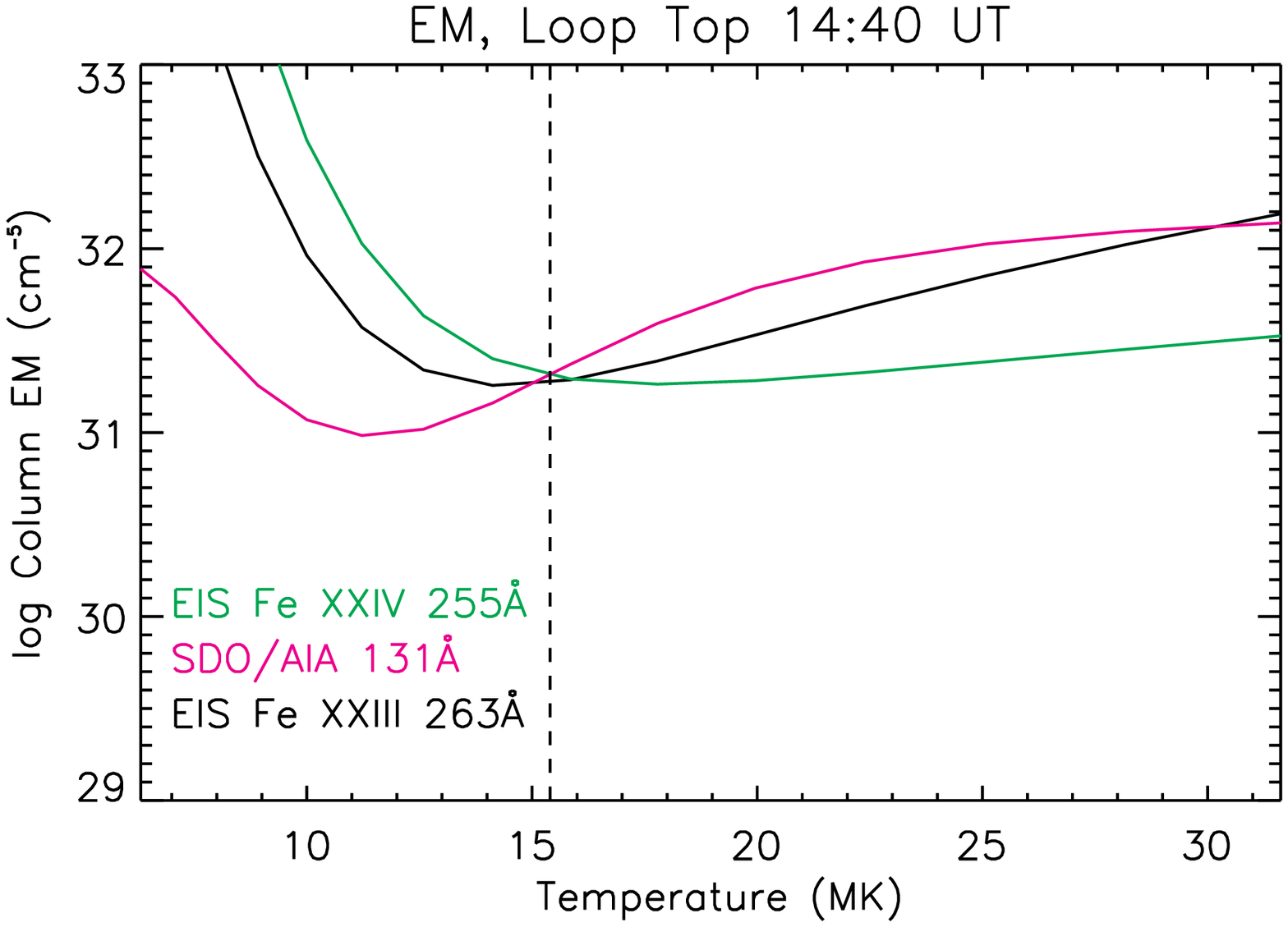}
	
\caption{Emission measure plot from EIS (\xxiii~263.77~\AA~and 255.11~\AA) spectral lines and AIA ~131\AA~observations at the loop top at the two different flare peak as observed by the GOES satellite $\sim$ 14:27 UT (top) and $\sim$ 14:40 UT (bottom).}
  \label{Fig:EM_plot}
  \end{figure}

%
Finally, we emphasize that the temperatures derived from spatially unresolved GOES (using coronal abundances, see \citet{White2005}) and RHESSI (Sect. \ref{Sect.:2.4}) data are consistent with the results from XRT:~$\approx$~24~MK (log($T$[K])~$\approx$~7.38) during the main impulsive phase (first and second rows in Fig. \ref{Fig:XRT_maps}) and 20~MK and 22~MK, from GOES and RHESSI respectively, at the second peak, around 14:40~UT.

\subsection{Emission measure diagnostics}
\label{Sect.:5.4}
If the plasma is isothermal at a temperature $T_0$, the emission measure of an optically thin spectral line can be expressed as:
\begin{equation}
EM_h(\lambda)=I_\mathrm{obs}(\lambda)/A(X)G(N_\mathrm{e}, T_0)
\label{eq:EM}
\end{equation}

where $I_\mathrm{obs}$ is the observed intensity, $A(X)$  is the element abundance relative to hydrogen, and $G(N_\mathrm{e}, T_0)$ is the contribution function of the line. For each line and temperature $T_\mathrm{i}$, the $I_\mathrm{obs}/A(X)G(N_\mathrm{e}, T_\mathrm{i})$ ratio represents an upper limit to the value of the emission measure at that temperature. For different spectral lines, the loci of the curves $I_\mathrm{obs}/G(N_\mathrm{e},T)$ at a given density $N_\mathrm{e}$ therefore provides an upper limit to the value of the emission measure of the plasma. 

In Fig. \ref{Fig:EM_plot}, we plot the ratio $I_\mathrm{obs}(\lambda)/G(N_\mathrm{e},T)$ for the EIS \xxiii~263.76~\AA~and \xxiv~255.11~\AA~spectral lines observed at the loop top of the flare loops at two times (14:27 and 14:40~UT) close to the two flare peaks observed in the soft X-ray light curves (Fig. \ref{Fig:goes}). 

The intensity $I(\lambda)$ of the EIS lines has been converted from data number (DN) to physical units (erg s$^{-1}$ sr$^{-1}$ cm$^{-2}$ $\AA^{-1}$), as described in the EIS Software note no.2 \footnotemark[4].  
The factor $G(N_e,T_0)$ in eq. \ref{eq:EM} is calculated by using the \emph{gofnt.pro} routine available within the CHIANTI package \citep{Dere:1997, Landi:2013}, assuming photospheric abundances. 

Moreover, we can estimate the column emission measure of the high temperature plasma from AIA observations in the 131~\AA~channel. For AIA observations, the $EM_h$ can be expressed as
\begin{equation}
EM_h=I_\mathrm{obs}(131~\AA)/R(N_\mathrm{e}, T_0)
\end{equation}
where $R(N_\mathrm{e}, T_0)$ is the AIA 131~\AA~temperature response. The response was calculated by using the AIA effective area and new atomic data from CHIANTI v 7.1 as described in \citet{DelZanna11b}. The count rates $I_\mathrm{obs}$(131~\AA)  have been averaged over the same location where we measured the intensity of the \xxiii~and \xxiv~spectral lines. 

The EM curves in the top panel of Fig. \ref{Fig:EM_plot} consistently show an isothermal loop structure at a temperature of about 18.5 MK at around 14:27 UT (first peak of the flare). By 14:40 UT, the loop temperature then drops to about 15.4 MK by 14:40 UT (second peak of the flare) as shown in the bottom panel of Fig. \ref{Fig:EM_plot}. These results are consistent (within an uncertainty of $\sim$~15\%) with the loop top temperatures observed at the same times by XRT ($\sim$ 18.7~MK and 13.5~MK, see Fig. \ref{Fig:XRT_maps}). 

The main source of uncertainty of the $EM_h$ values is given by the absolute calibration uncertainty of EIS and AIA. The radiometric calibration of the EIS spectrometer has been revised \citet{DelZanna13} and \citet{Warren14} to take into account the degradation of the instrument's performance over time. In this work, we applied the radiometric calibration factors found by \citet{DelZanna13}. However, we have verified that (within the uncertainty) the corrections introduced by the two methods give similar results for the EIS spectral lines used in this work. For instance, using the radiometric calibration by \citet{Warren14} would result in a difference of $\sim$ 2~MK in the loop-top temperatures, which is well within the errors on the calibrated intensities estimated by the two methods. 
In addition, an uncertainty of $\sim$ 25 $\%$ can be assumed for the photometric calibration of SDO/AIA \citep{Boerner12}. Therefore, we assume a 25 $\%$ uncertainty for the estimates of $EM_h$ values.

With some assumptions, the EM analysis also allows us to provide a lower limit value of the flare plasma density.
The column emission measure (EM) is in fact defined as  
\begin{equation}
EM_h= \int_h N_\mathrm{e} N_\mathrm{H} dh
\label{Eq:EM1}
\end{equation}
 which can be approximated by 
 \begin{equation}
 EM_h \simeq 0.83 \langle N_\mathrm{e} \rangle^2 \Delta h
 \label{Eq:EM2}
\end{equation}
   where $\Delta h $ is the column depth of the emission layer and $N_\mathrm{H} =0.83 N_\mathrm{e}$ in a fully ionized gas with helium abundance relative to hydrogen $A(He)=0.1$. In addition, we have assumed a \emph{spectroscopic filling factor} equal to 1. Assuming that the loops have a circular shape, the depth $\Delta h$ can be estimated from the width of the loop structures visible in the AIA 131~\AA~images. We obtained widths of $\sim$~3~$\cdot$ 10$^{8}$~cm. 
By using the eq. \ref{Eq:EM2}, we then obtained electron densities $N_\mathrm{e}$ at the top loop of $\sim 8 \cdot 10^{10}$ cm$^{-3}$ and $\sim 1 \cdot 10^{11}$ cm$^{-3}$ at the times $\sim$ 14:27~UT and 14:40~UT, respectively. It is interesting to note that the density at the loop apex remains almost constant between the two peaks. This might suggest that when the second event occur, the hot loops were already dense and pre-filled by flare plasma from the previous chromospheric evaporation event.

Moreover, it is important to point out that if one assumes coronal abundances (e.g. with Fe abundance increased by a factor 4 from the photospheric value \citep{Feldman92}), the densities become smaller by a factor $\sim$ 2. Finally, we emphasize that these density values represent a lower limit due to the assumption of a filling factor equal to 1.

 \footnotetext[4]{http://solarb.mssl.ucl.ac.uk:8080/eiswiki/}

 \section{Comparison with hydrodynamics simulations with HYDRAD}
\label{Sect.:6}
In order to simulate the response of the plasma to the heating during the flare, we have run 1D hydrodynamics simulations with the HYDRAD code \citep{bradshaw2003,bradshaw2013,reep2013}. This section is organized as follows: Sect. \ref{Sect.:6.1} describes the details of the numerical experiments, Sect. \ref{Sect.:6.2} shows the results of the simulations, which are then compared with the observational results in Sect. \ref{Sect.:6.3}. 

\subsection{Modeling}
\label{Sect.:6.1}
The HYDRAD code solves the hydrodynamic equations of conservation of mass, momentum, and energy for an isolated magnetic flux tube with a multi-fluid plasma (electrons, ions, and neutrals).  The equations and assumptions are detailed in the appendix of \citet{bradshaw2013}.  The loops are assumed to be semi-circular, along the field-aligned direction, with a constant cross-sectional area.  The initial temperature and density profiles are found by integrating the hydrostatic equations from the chromosphere to the apex of the coronal loop.  The electron and ion populations are assumed to be in thermal equilibrium before any significant heating occurs.  We estimate the length of the loop from the foot-point separation measured in AIA images.  
We adopt an electron beam heating model, whereby electrons accelerated to high energies near the top of the loop stream from the corona to the chromosphere, depositing their energy through Coulomb collisions with the ambient plasma \citep{brown1971, emslie1978}.  We assume an electron distribution with a sharp cut-off \citep{holman2011}, that is, no electrons at energies beneath the low-energy cut-off $E_{c}$.  The heating deposition is given by \citet{emslie1978}, although the expressions have been modified based on the ideas of \citet{hawley1994}, which generalizes the result to a non-uniform ionization structure, which is important for recovering breaks in the photon spectra \citep{kontar2011}.  



The time profile of this event is taken by normalizing the derivative of the GOES 1--8 \AA\ light-curve.  In other words, we assume that the Neupert effect \citep{Neupert1968,dennis1993} is valid for this flare, namely that the hard X-ray flux (and thus heating rate by electrons) is proportional to the time derivative of the soft X-ray flux. We thus use the time derivative of the GOES light-curve, which is indeed similar to the 25--50 keV curve from RHESSI, starting at 14:00 UT and lasting until the derivative becomes negative at around 14:50 UT.  We estimate from the RHESSI data that the low-energy cut-off is approximately 20 keV.  

\begin{figure}
\begin{minipage}[b]{0.5\linewidth}
\centering
\includegraphics[width=3.5in]{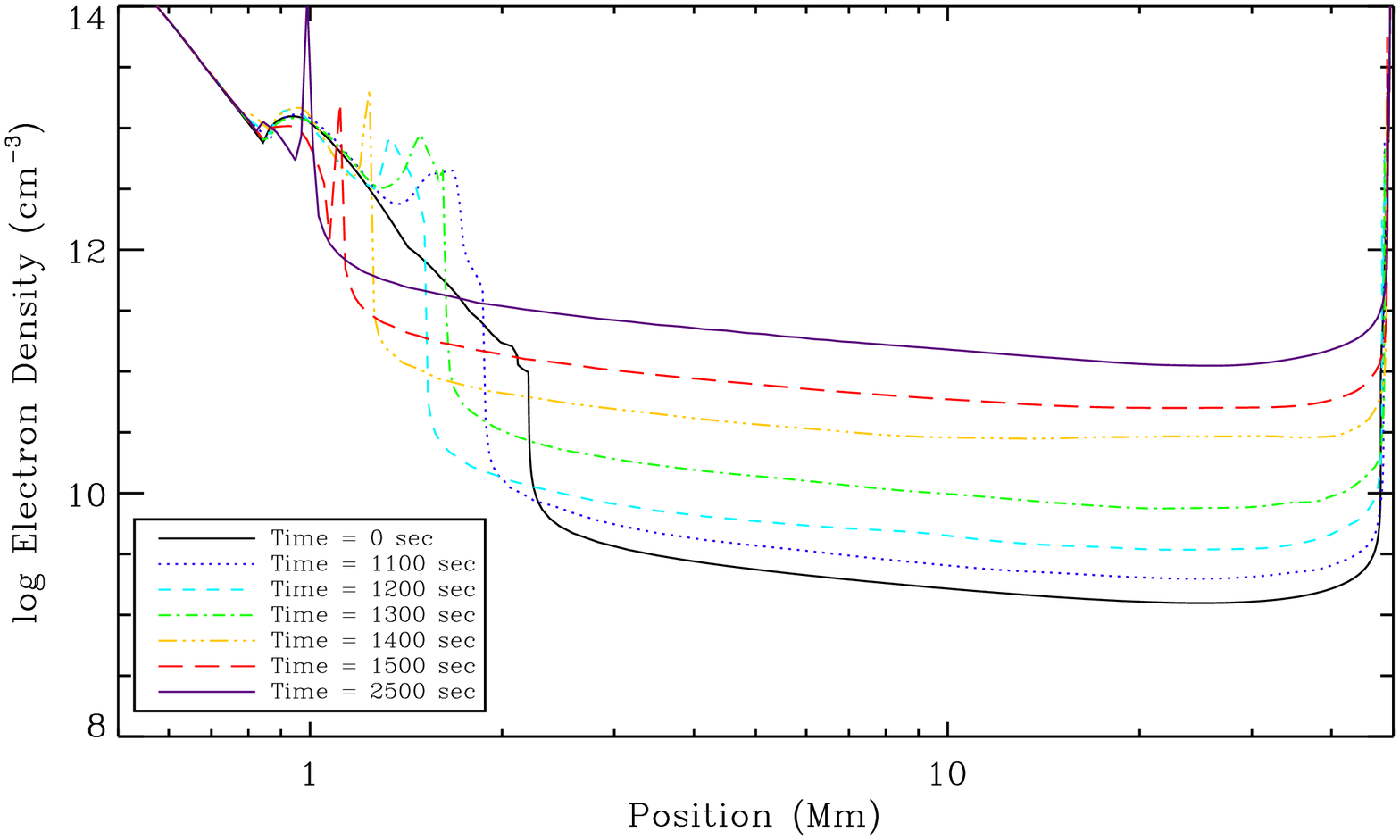}
\end{minipage}
\hspace{0.1in}
\begin{minipage}[b]{0.5\linewidth}
\centering
\includegraphics[width=3.5in]{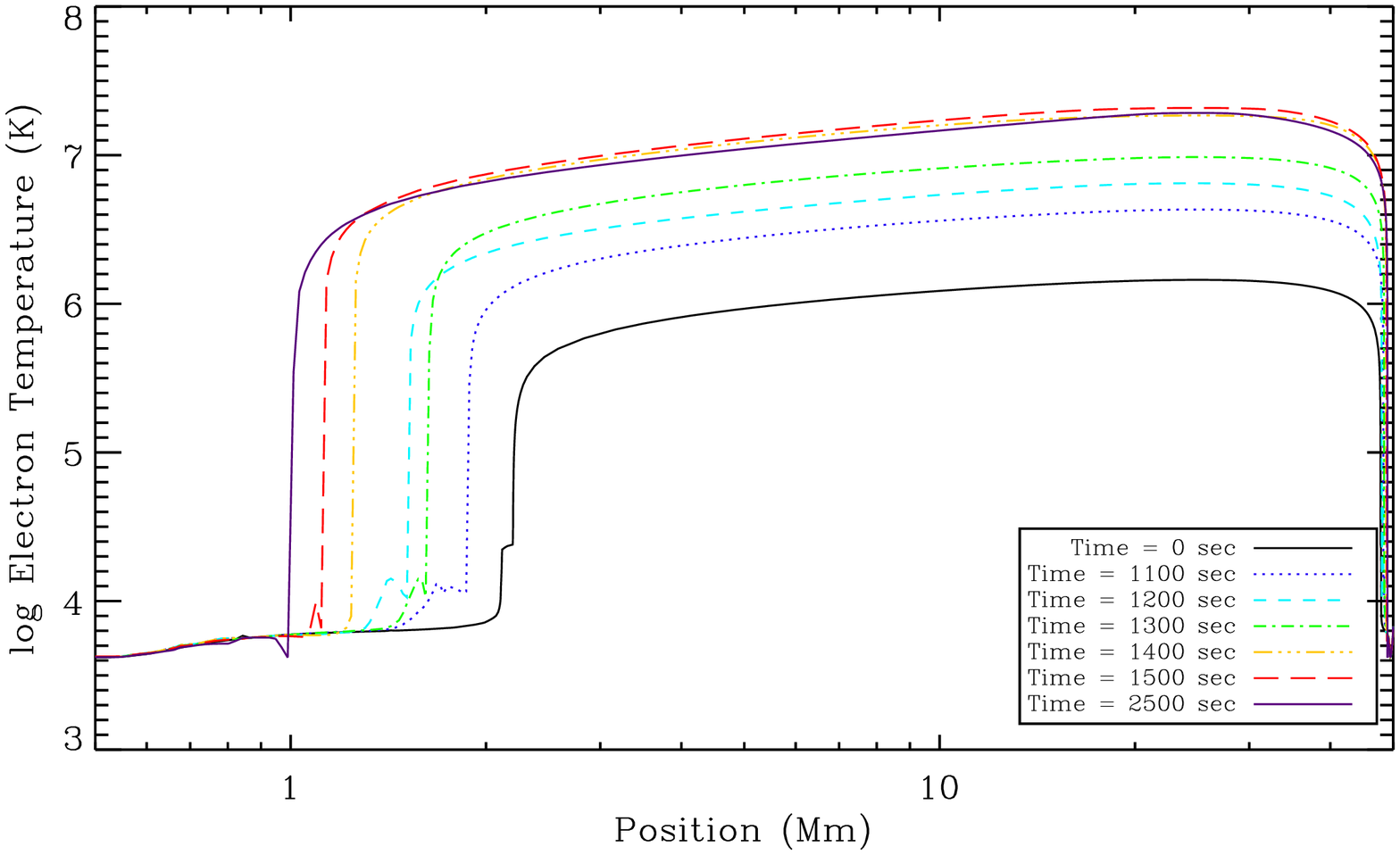}
\end{minipage}
\begin{minipage}[b]{0.5\linewidth}
\centering
\includegraphics[width=3.5in]{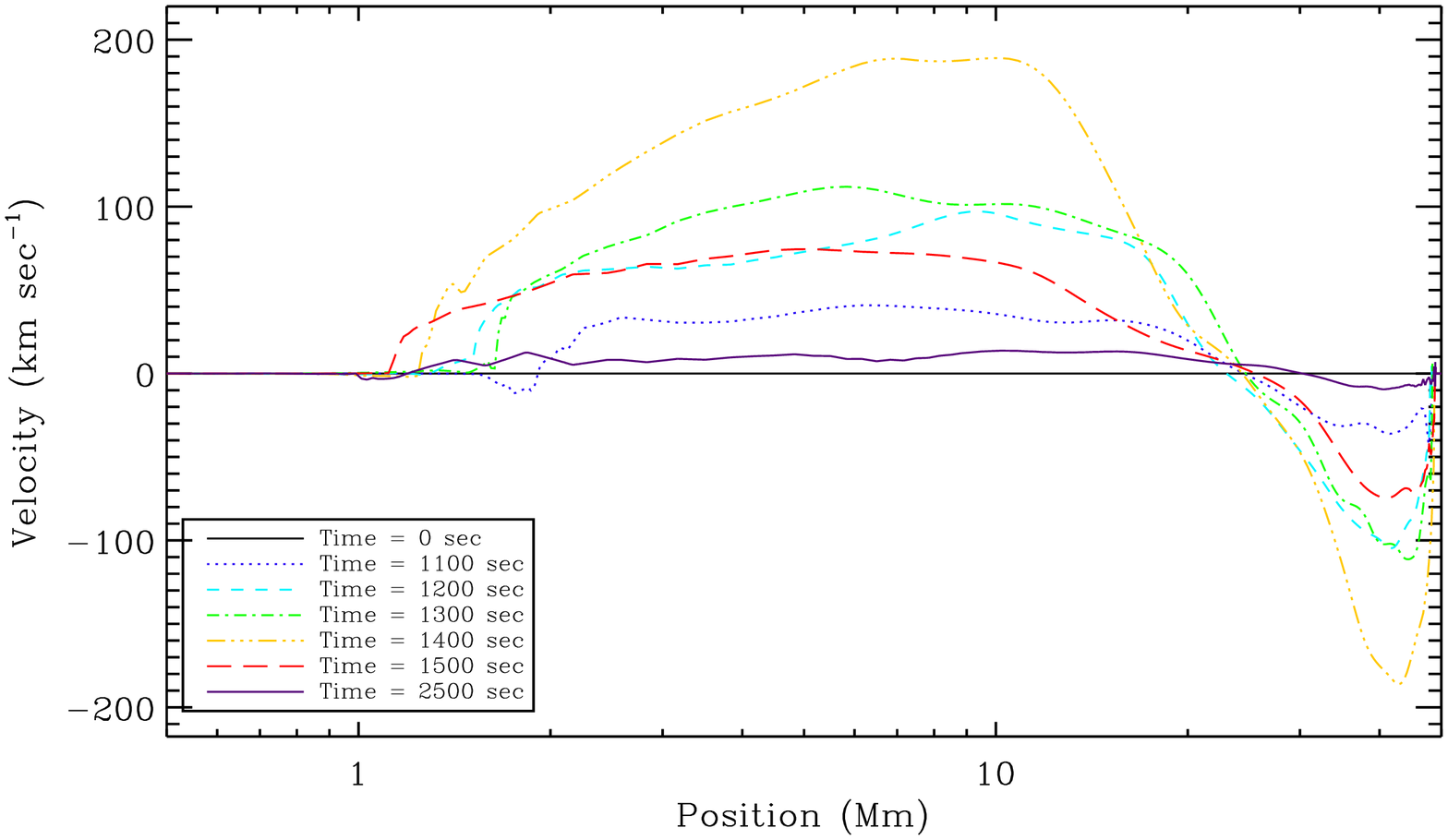}
\end{minipage}
\hspace{0.1in}
\caption{The atmospheric response of the simulation.  At the top panel, the (electron) density vs position, with a few times overplotted and in the middle panel, the (electron) temperature vs position.  At bottom, the bulk flow velocity vs position, where right-moving flows are defined as positive.}
\label{Fig:atmresponse}
\end{figure}
 
\begin{figure}
\begin{minipage}[b]{0.5\linewidth}
\centering
\includegraphics[width=3.5in]{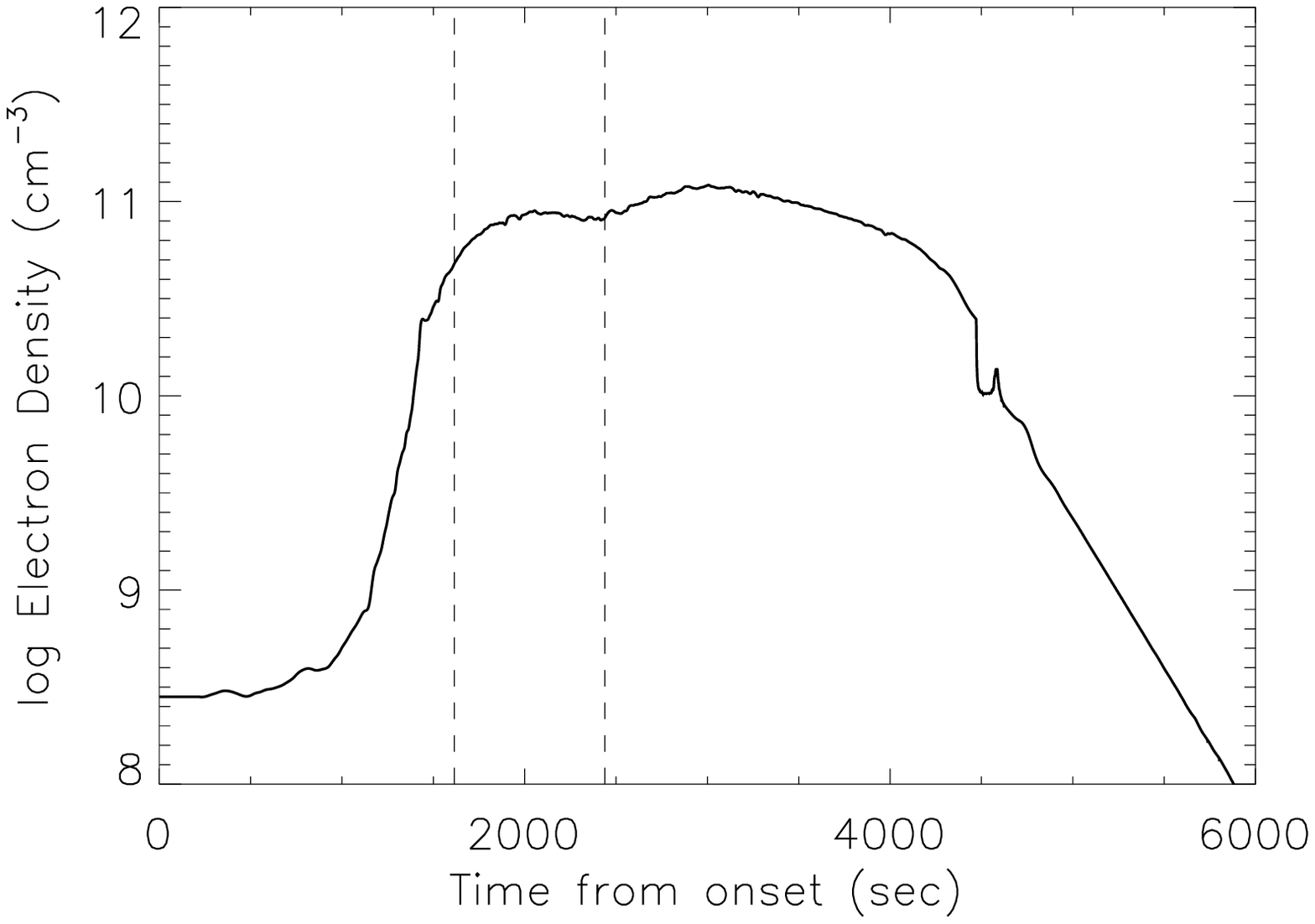}
\end{minipage}
\hspace{0.1in}
\begin{minipage}[b]{0.5\linewidth}
\centering
\includegraphics[width=3.5in]{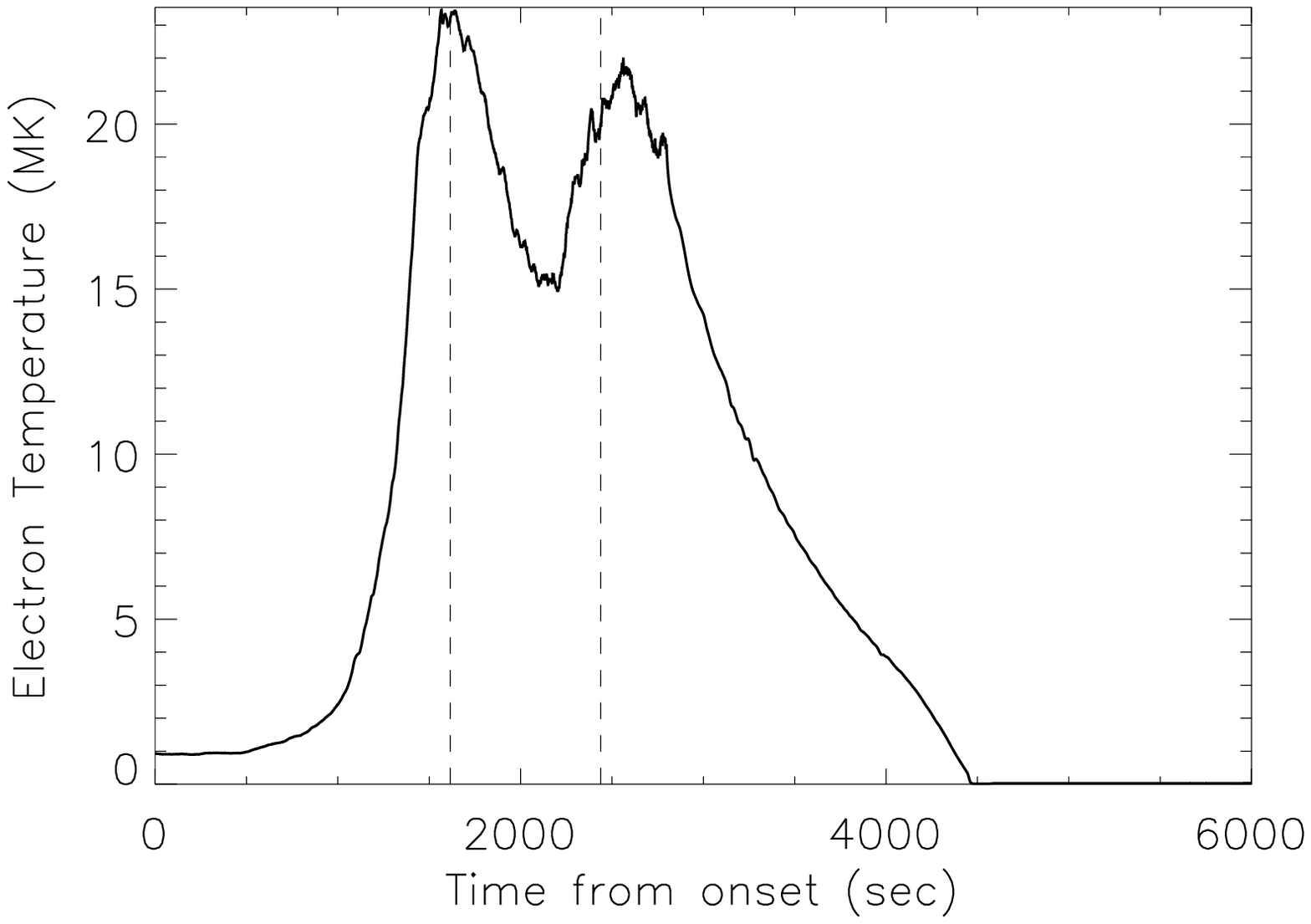}
\end{minipage}
\caption{The values of the (electron) density (top) and temperature (bottom) at the apex of the coronal loop as functions of time.  The vertical lines indicate the approximate time when the observational values were determined.  }
\label{Fig:apex}
\end{figure}

\subsection{Simulation Results}
\label{Sect.:6.2}
\begin{figure}
\begin{minipage}[b]{0.5\linewidth}
\centering
\includegraphics[width=3.5in]{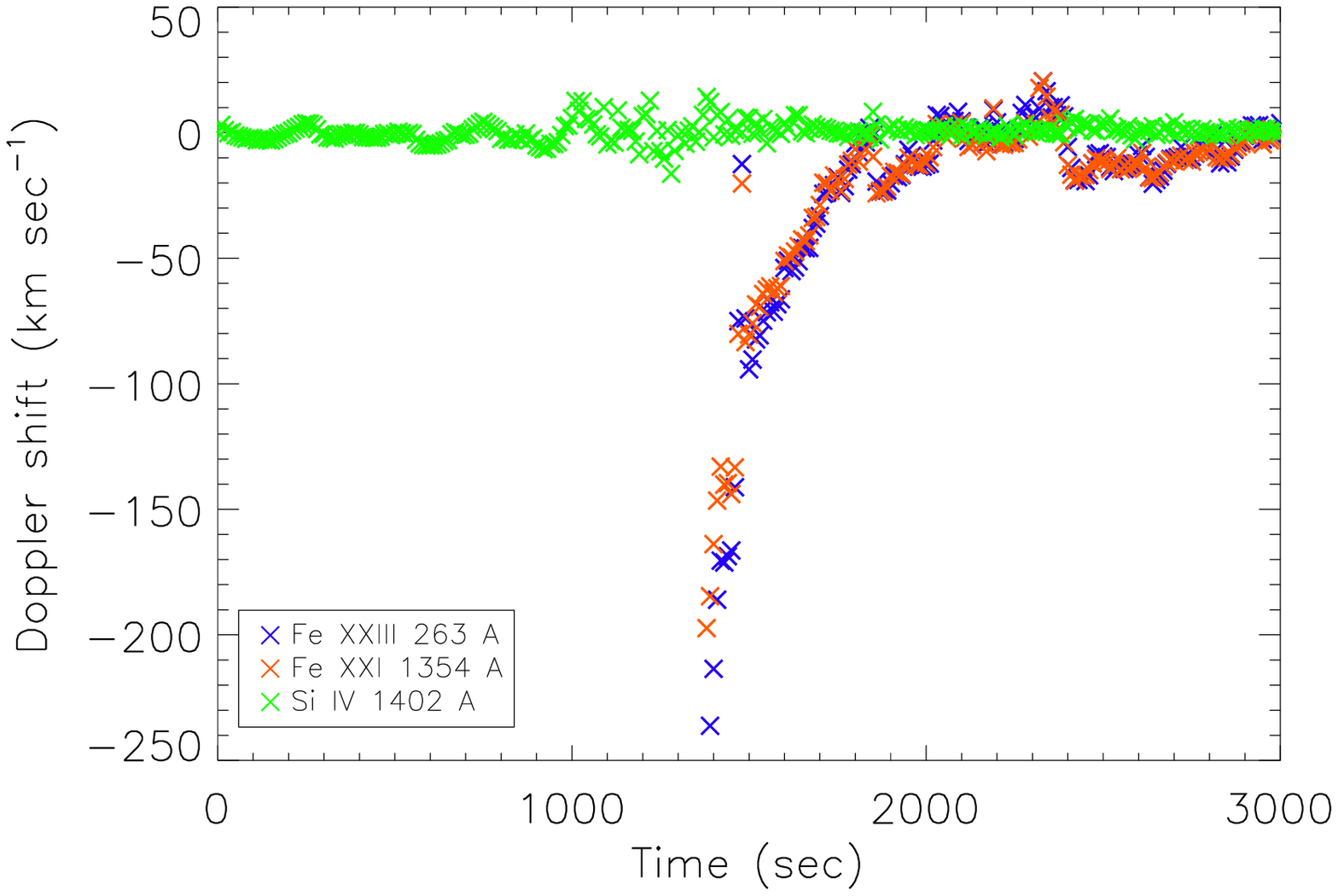}
\end{minipage}
\hspace{0.1in}
\begin{minipage}[b]{0.5\linewidth}
\centering
\includegraphics[width=3.5in]{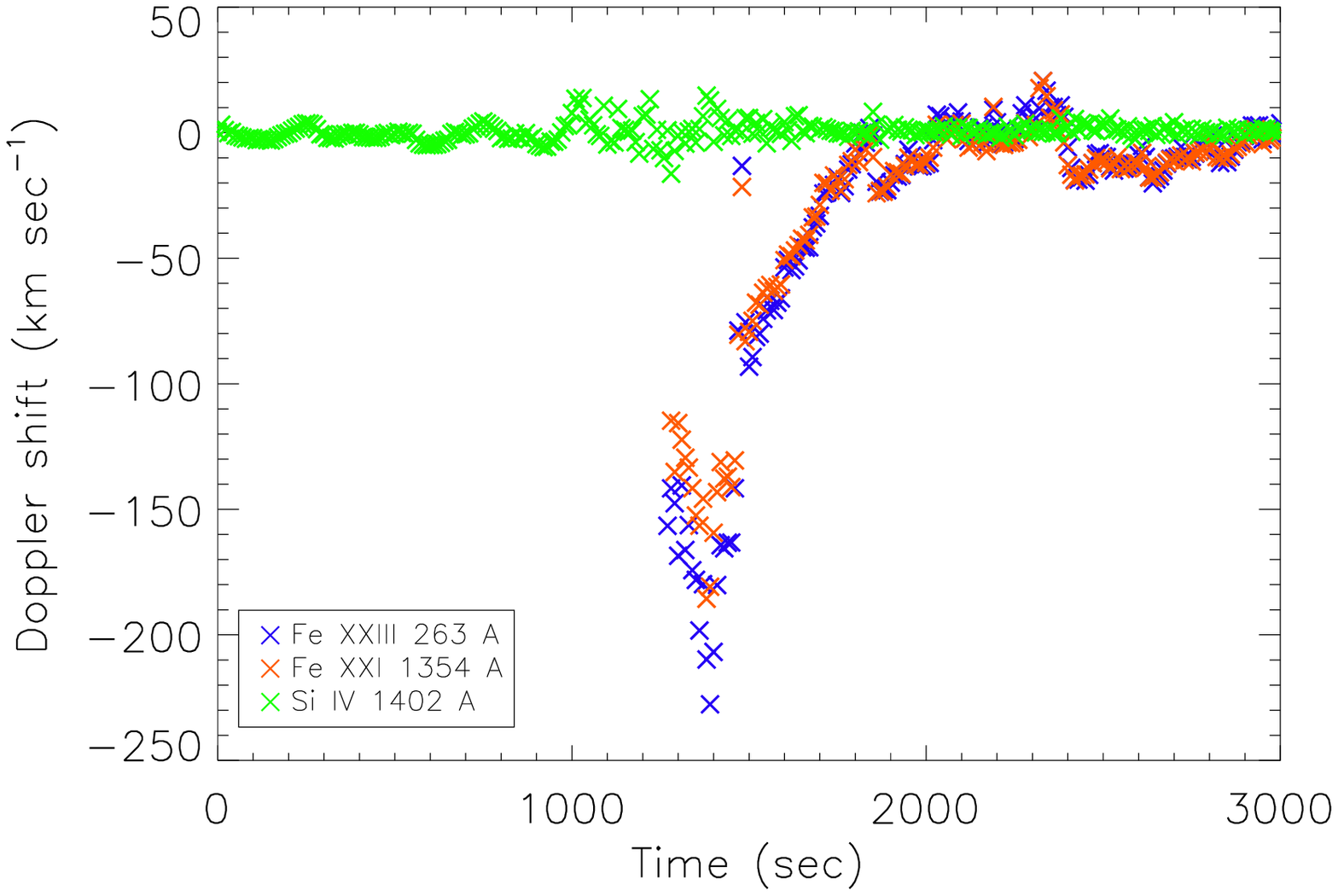}
\end{minipage}
\caption{Synthesized Doppler shifts for 3 spectral lines as might be observed with EIS and IRIS, as functions of time.  The top panel shows calculations with full non-equilibrium ionization, while the bottom one assumes equilibrium ionization.}
\label{Fig:shifts}
\end{figure}

The atmospheric responses to the heating are shown in Fig. \ref{Fig:atmresponse}, showing the electron density, electron temperature, bulk flow velocity, and energy deposition as functions of position, for a few selected times.  The majority of the heating begins around 1000 seconds into the simulation, when the temperature begins to rapidly rise from an initial coronal temperature of less than 1 MK to well over 20 MK.  As the chromosphere is heated up to chromospheric values, chromospheric evaporation begins to cause strong flows of material into the corona.  These flows quickly raise the electron density by more than two orders of magnitude in the corona, to above $10^{11}$ cm$^{-3}$.
Fig. \ref{Fig:apex} shows the electron density and temperature at the apex of the loop as functions of time.  The temperature has two distinct peaks, one around 22 MK and the other around 19 MK, corresponding to the two heating events seen in the X-ray light-curves.  The density increases to above $10^{11}$ cm$^{-3}$, and remains roughly constant during the duration of the flare, until heating ceases and the loop begins to cool and drain at late time.  After approximately 4000 seconds, the loop cools catastrophically \citep{cargill2013}.  
From this simulation, three spectral lines were forward modeled using the method of \citet{bradshaw2011} with atomic data from CHIANTI v7.1: Fe XXIII 263.77~\AA~as might be seen by EIS, Fe XXI 1354.08~\AA\, and Si IV 1402.77~\AA~as might be seen by IRIS.  The lines were fitted with Gaussians at every point in time, from which Doppler shifts were measured.  Fig. \ref{Fig:shifts} shows synthesized Doppler shifts for the 3 spectral, which can be compared directly with the observed quantities in Fig \ref{Fig:Line_profile}.  We emphasize that the basic duration of evaporation and magnitude of the flows are consistent with the observations.  

 \subsection{Comparison of the derived plasma parameters with those in the HYDRAD simulation}
 \label{Sect.:6.3}
In this section we compare the plasma parameters derived from Sects. \ref{Sect.:4} and \ref{Sect.:5} with the outcomes of the HYDRAD 1D hydrodynamic simulations of a flare loop heated by an electron beam.\\
\begin{enumerate}[(a)]

\item The \xxi~1354.08~\AA~and \xxiii~263.77~\AA~velocities shown in Fig. \ref{Fig:Line_profile} are consistent with the HYDRAD synthesized Doppler shifts reported in Fig. \ref{Fig:shifts}. In particular, we observe upflows velocities of up to $\sim$ 200 km\,s$^{-1}$ in the \xxi~line with IRIS, in agreement with the simulated results. \\
Moreover, simultaneous \xxiii~observations with EIS show higher velocity values than the \xxi~line, in accordance with the model prediction of a linear relationship between the Doppler shift and the temperature of formation, and previous observational results from CDS and EIS \citep[see for example][]{DelZanna06,Milligan09}.

Interestingly, we observe that the \xxi~line is first blue-shifted by $\sim$~ 160~km~s$^{-1}$ (at 14:18:37~UT) and then the evaporation further increases to $\sim$~200~km~s$^{-1}$ before gradually decreasing, as shown in Fig. \ref{Fig:Line_profile}. This trend would suggest that our results are in agreement with the HYDRAD simulation obtained assuming ionization equilibrium (Fig. \ref{Fig:shifts}, top panel). However, as explained in Sect. \ref{Sect.:4}, we are not able to observe a \xxi~profile before 14:18:37~UT, due to the strong FUV continuum dominating the \oi~spectral window. Hence, we cannot definitively confirm whether the \xxi~blueshifts trend is consistent with the ionization equilibrium rather than the non-equilibrium simulation. 

We note that the time of the maximum upflows in the simulation is predicted at $\sim$~1400~s after the start of the heating at 14:00~UT (Fig.\ref{Fig:shifts}), while the maximum \xxi~and \xxiii~blueshifts are observed at around 1100--1200~s after 14:00~UT (Fig.\ref{Fig:Line_profile}). Nevertheless, the amplitude of the flows and the trends are in very good agreement, and this is particularly pleasing considering that we are modelling a complex flare geometry with a single one-dimensional loop.

\item The \siiv~1402.77~\AA~sythesized velocities in Fig.\ref{Fig:Line_profile} (green) show that the transition region line is redshifted by less than 20~km\,s$^{-1}$ during the impulsive phase. This is consistent with the IRIS observation that the only significant \siiv~redshift is $\sim$ 16~km\,s$^{-1}$ at around 14:18~UT (see Fig. \ref{Fig:Line_profile}).
\item  By using the ratio of the IRIS \oiv~lines (formed at log($T$[K])~$\sim$~5.2) we measured electron densities equal to or greater than $10^{12}$ cm$^{-3}$ at the flare footpoints between~$\sim$~14:17 and 14:23~UT.~In addition, the EIS \xiv~lines ratio gives an electron density for the log($T$[K])~$\sim$~6.3 plasma at the ribbon which increases from $\sim$ $10^{10}$ cm$^{-3}$ to $10^{11}$ cm$^{-3}$ from around 14:16~UT to 14:20~UT, and then it remains almost constant from 14:20~UT onward.

We can compare these observational values with the results of the simulated density (density vs loop position, on a logarithmic scale) shown in the first panel of Fig. \ref{Fig:atmresponse} for different time intervals. The time is indicated in seconds from 14:00~UT, which is the time of the start of the heating in the simulation. 

Fig.\ref{Fig:atmresponse} shows that the electron density of the flaring plasma at the flare footpoint is within the range $\sim$ $10^{10}$ to $10^{12.5-13}$ cm$^{-3}$ during the interval 1100--1200 s (corresponding to $\sim$ 14:18--14:20~UT) and then it rises to $10^{10.5}$--$10^{13}$ cm$^{-3}$ 1300 s after the start of the heating (corresponding to $\sim$~14:22~UT). The footpoint density then increases to $10^{11}$--$10^{13}$ cm$^{-3}$ at the time 1400 s in the simulation ($\sim$~14:23~UT). Finally, it remains within the range $10^{11.3}$--$10^{13}$ cm$^{-3}$ from 1500 s ($\sim$ 14:25~UT) onward. 

The observational results are therefore well in agreement with the range of density values obtained with HYDRAD, bearing also in mind that the density sensitivity of both the \oiv~and the \xiv~ratio is limited by a high density value and therefore densities above this limit cannot be calculated. In addition, we would like to point out that the density of the \xiv~plasma represents a lower estimation given the uncertainty associated with the \sivii~blending with the \xiv~274.20~\AA~line, as explained in Sect. \ref{Sect.:5.2}.

\item The second panel of Fig.\ref{Fig:atmresponse} shows the time evolution of the simulated electron temperature along the loop. In particular, we observe that the temperature of the flare loops increases to up to $\sim$ log($T$[K])~$\sim$~7.2--7.4 after $\sim$ 1400--1500 s ($\sim$ 14:23--14:25~UT) and remains almost constant at 2500 s (corresponding to $\sim$ 14:40~UT). This is in agreement with what is shown in the AIA and XRT temperature maps of Figs. \ref{Fig:temp_maps} and \ref{Fig:XRT_maps}, where the strong hot emission at log($T$[K])~$>$~7.20 is observed to dominate the flare loops after $\sim$~14:23~UT. 

\item An important confirmation of our model is obtained from comparing the apex temperatures and densities at the two flare peaks in the hydrodynamic model with the observed values obtained with AIA and EIS (see Fig. \ref{Fig:EM_plot}). The EM loci curves in the left panels of Fig. \ref{Fig:EM_plot} consistently show that the plasma temperature at the loop top decreases from $\sim$~18.5~MK to $\sim$~15~MK from 14:27~UT to 14:40~UT (also consistent with the XRT temperature measurements). We emphasize that these values are close to the temperatures observed by XRT (Fig. \ref{Fig:XRT_maps}). Correspondingly, the electron density remains almost constant, varying from $\sim$~8~$\cdot$ $10^{10}$ cm$^{-3}$ to $10^{11}$ cm$^{-3}$. However, we note that the densities derived from the EM and estimate of loop size represent lower limits due to the assumption of a spectroscopic filling factor equal to 1. 

The observed values of peak temperature and density at the loop top are in good agreement with the curves reported in the right panels of Fig. \ref{Fig:EM_plot}, showing the apex density and temperature over time as simulated by HYDRAD. The horizontal dotted lines in the figure indicate the times of the EM loci results ($\sim$ 14:27~UT and 14:40~UT). In particular, the observed values of density and temperature at these two times agree within a 20 $\%$ uncertainty with the simulated ones. In addition, the simulation confirms the observational evidence that the apex temperature decreases by approximatively 3--4~MK between the two peaks, while the electron density remains almost constant.  The fact that we observe a steady high density in the loop over several minutes might be explained as the result of the two consecutive heating events, which allowed the flare loops to be continuously replenished by hot plasma material.  

\item We simulated an asymmetrically heated loop in order to reproduce the asymmetric high temperature structures that we observe in the AIA~131~\AA~and EIS \xxiii~images (see Fig. \ref{Fig:overview_131_SJI}, \ref{Fig:overview_EIS} and the online movies 1, 2). The results of density, temperature and Doppler shift values remain similar to the case of a symmetric heating. However, the maximum velocity and density of the hot plasma upflows are located higher up in the loop, i.e., in a higher detector pixel than the cool ribbon emission. This might explain why the \xxi~evaporating plasma is observed few pixels above the FUV continuum and the cool emission lines from the ribbon, as observed in the present and other flare studies with IRIS \citep{Young15}.

\item The observations of the HXR burst with RHESSI suggest a strongly non-thermal component, with photon energies extending well above 25~keV, pointing to the likelihood of non-thermal accelerated electrons with energies higher than that.  It seems unlikely that a purely thermal model could simultaneously reproduce the high energy emission while maintaining a temperature close to that which is measured with the spectroscopic and imaging instrumentation. In particular, the strong HXR burst observed with RHESSI at the same time as the plasma upflows would require a temperature that far exceeds the values found by combined diagnostics from AIA, XRT and EIS. 
Finally, we would like to point out that microwave emission was detected by the Radio Solar Telescope Network (RSTN) during this flare. The observation of microwave emission during flares provides a direct signature of non-thermal electrons \citep{Bastian1998,White2011}. The shape and intensity of the microwave spectrum is in fact a clear evidence of gyrosynchrotron emission by non-thermal electrons.

\item The cut off energy and spectral index of the electron beam used in the simulations are consistent to the values obtained from RHESSI. However, using the energy flux input as derived by RHESSI would result in very high values of plasma temperatures and densities which would not been consistent with those derived from the AIA, EIS and IRIS observations and typically observed during flares. Therefore, we used an energy input which is a factor of 10 lower in our simulations. One possible explanation for this discrepancy might be related to the estimation of the footpoint area, which is quite complicated given the complexity of the flare. In addition, most of the HXR emission seems to originate from the loop, which implies a high density there. Because of this, most of the energy might be deposited in the loops, and therefore it is likely that it did not reach the chromosphere. In addition, it is important to note that our simulations include a single flare loop, while the energy flux might have been distributed over a sequence of loops.
 
\end{enumerate}

\section{Summary and conclusions}
\label{Sect.:7}
In this work we have presented an extensive study of an X-class flare which occurred in the AR 12192 on the 27 October 2014. This flare was observed with IRIS, Hinode,  AIA and RHESSI. 

In this study, we are able to combine simultaneous and high cadence observations of the \xxi~1354.08~\AA~and \xxiii~{263.77~\AA~}high temperature emission at the flare kernels during the chromospheric evaporation phase. We believe this to be the first such analysis of this nature. Our aim has been to investigate the response of the chromospheric plasma to the heating, and to compare the results with the predictions of theoretical models at two different (but similar) temperatures. In addition, we are interested in understanding to what extent the instrumental resolution represents a limitation when studying the dynamics of the flaring plasma. 

Different plasma parameters (Doppler shifts, density and temperature) derived from combined IRIS/Hinode/AIA observations have been compared to a detailed model of a flare loop undergoing heating by an electron beam with the 1D hydrodynamics HYDRAD code. 

The main results of our work are summarized as follows:  

\begin{enumerate}
\item The site of high temperature upflows from the flare kernels seems to be resolved by IRIS, which observes completely blue-shifted single \xxi~line profiles of up to $\sim$ 200 km\,s$^{-1}$ with a raster cadence of 26~s during the rising phase of the flare. Taking into account the possible inclination of the loop, we suggest that the actual velocity could be higher.

\item At the same footpoint region, we observe \xxiii~blueshifts in two consecutive EIS rasters (with a 212 s cadence) during the peak of the chromosperic upflows. The \xxiii~line profiles are often asymmetric and were fitted with two Gaussian components, in contrast to the simultaneous IRIS \xxi~observations. We interpret this difference as due to the lower spatial resolution of EIS.

\item The dominant \xxiii~blue-shifted component shows a larger Doppler shift velocity value than the \xxi~line which is indicative of a faster evaporation at higher temperatures, in accordance with the hydrodynamical model. 

\item The values and the trend of the Doppler shift velocities with time are consistent with the results of the HYDRAD 1D hydrodynamics simulations of a flare loop where the heating is driven by a beam of accelerated electrons. 

\item Simulating an asymmetrically heated flare loop results in the observation of the high temperature upflows higher up along the loop, corresponding to few pixels above the position where we observe the cool temperature lines in the IRIS detector. 

\item Electron density estimates from the \oiv~399.77~\AA~and 1401.16~\AA~line ratio indicate values equal or larger than $10^{12}$ cm$^{-3}$ at the flare ribbon during the impulsive phase. Interestingly, we found some unidentified cool emission lines in the spectral range $\sim$~1399--1406~\AA~which blend with the \oiv~1401.16~\AA~and 1399.77~\AA~lines. Their emission is enhanced during the impulsive phase of the flare and these need to be taken into account when performing any plasma diagnostics.

\item The EIS \xiv~line ratio gives electron density values of $10^{10}$  cm$^{-3}$ at the flare ribbon at the very beginning of the rise phase which then increase up to $10^{11}$  cm$^{-3}$ going toward the peak of the flare. 

\item Temperature estimates from the ratio of the AIA 131~\AA~and 94~\AA~bandpass and XRT filters show that high temperature emission (log($T$[K])~$>$ 7.20) is first concentrated at the flare PR ribbon where the high temperature \xxiii~emission originates and is then observed to fill the flare loops, in agreement with \citep{Simoes2015}.

\item The observed values of temperature and density at the flare ribbons as a function of time are consistent to the results of the HYDRAD simulations detailed in Sect.\ref{Sect.:6}.

\end{enumerate}

Finally, we would like to point out that while there are no a priori reasons to assume only two flow components for the fitting of the asymmetric line profiles with EIS, this approach provides a simple baseline against which we can interpret the observed spectra. Future work will include generalizing the fitting of IRIS and EIS flare lines to a continuum of Gaussian components by using the velocity differential emission measure analysis \citep[VDEM;][]{Newton95}. The VDEM method, combined with the adoption of a multi-threaded loop model in the 1D hydrodynamic simulations, will allow us to investigate further the validity of single or two-component Gaussian fit approach.

Simultaneous observations with different EUV, UV and X-ray spectrometers are crucial in order to investigate the plasma response to the heating and to differentiate between different heating mechanisms responsible for flares. In the present study, different observational results combined with modelling indicate that the physics of the flare is consistent with an electron beam heating scenario. However, in order to understand how well the observations can be reproduced by theory, we need to observe multiple plasma parameters with high resolution instruments. The IRIS instrument now allows us to study the details of the chromospheric response to the heating with unprecedented spatial and temporal resolution. Due to the longer cadence of the EIS study used for this flare($\sim$ 212 s), we were able to combine the study of the \xxi~and \xxiii~plasma at only two particular times during the peak of the chromospheric evaporation. For future coordinated observations, we recommend higher cadence observations at the same footpoint regions as IRIS. In addition, statistical studies combining observations with modelling would be essential in order to investigate if a common mechanism may be responsible for the production of flares at different scales.

\begin{acknowledgements}
VP acknowledges support from the Isaac Newton Studentship, the Cambridge Trust, the IRIS team at Harvard-Smithsonian Centre for Astrophysics and the RS Newton Alumni Programme. This research was performed while JWR held an NRC Research
Associateship award at the US Naval Research Laboratory with support from NASA, and previously a PDRA at the University of Cambridge. HEM and GDZ acknowledge support from the STFC and the RS Newton Alumni Programme. LG and KR are supported by contract 8100002705 from Lockheed-Martin to SAO.PJAS acknowledges support from the European Community's Seventh Framework Programme (FP7/2007-2013) under grant agreement no. 606862 (F-CHROMA). JD acknowledges support from the RS Newton Alumni Programme.

 IRIS is a NASA small explorer mission developed and operated by LMSAL with mission operations executed at NASA Ames Research center and major contributions to downlink communications funded by the Norwegian Space Center (NSC, Norway) through an ESA PRODEX contract. Hinode is a Japanese mission developed and launched by ISAS/JAXA, with NAOJ as domestic partner and NASA and STFC (UK) as international partners. It is operated by these agencies in co-operation with ESA and NSC (Norway). AIA data are courtesy of NASA/SDO and the respective science teams. CHIANTI is a collaborative project involving researchers at the universities of Cambridge (UK), George Mason and Michigan (USA). 

\end{acknowledgements}

 \bibliographystyle{apj}
\bibliography{v8}

\end{document}